\documentclass[a4paper,11pt]{article}
\pdfoutput=1 
\usepackage{jheppub, bm, color} 
\usepackage[T1]{fontenc}
\usepackage{amssymb,amsfonts,slashed,amsthm,amsmath,graphicx, soul}
\title{Hessian eigenvalue distribution in a random Gaussian landscape}

\author{Masaki Yamada,}
\author{Alexander Vilenkin}

\affiliation{Institute of Cosmology, Department of Physics and Astronomy, 
Tufts University, Medford, MA  02155, USA}

\emailAdd{Masaki.Yamada@tufts.edu}
\emailAdd{vilenkin@cosmos.phy.tufts.edu}

\def\beqa{\begin{eqnarray}}
\def\eeqa{\end{eqnarray}}

\def\nn{\nonumber}

\def\rmk{\right)}
\def\lmk{\left(}

\def\({\left(}
\def\){\right)}
\def\[{\left[}
\def\]{\right]}
\def\det{{\rm det}}
\def\nn{\nonumber \\}

\def\pot{U}
\def\field{\phi}
\def\hs{\zeta}

\def\Tr{{\rm Tr}}

\def\lmk{\left(}
\def\rmk{\right)}
\def\lkk{\left[}
\def\rkk{\right]}
\def\dd{{\rm d}}
\def\la{\left<}
\def\ra{\right>}
\newcommand{\eq}[1]{Eq.~(\ref{#1})}

\newcommand{\beq}{\begin{eqnarray}} 
\newcommand{\eeq}{\end{eqnarray}}

\newcommand{\bel}[1] {\begin{equation}\label{#1}}
\newcommand{\beal}[1] {\begin{eqnarray}\label{#1}}
\newcommand{\be}{\begin{equation}}
\newcommand{\ee}{\end{equation}}
\newcommand{\bea}{\begin{eqnarray}} 
\newcommand{\eea}{\end{eqnarray}}
\newcommand{\abs}[1]{\lvert#1\rvert}

\def\del{\partial}

\abstract{
The energy landscape of multiverse cosmology is often modeled by a multi-dimensional random Gaussian potential.  The physical predictions of such models crucially depend on the eigenvalue distribution of the Hessian matrix at potential minima.  In particular, the stability of vacua and the dynamics of slow-roll inflation are sensitive to the magnitude of the smallest eigenvalues.  The Hessian eigenvalue distribution has been studied earlier, using the saddle point approximation, in the leading order of $1/N$ expansion, where $N$ is the dimensionality of the landscape.  This approximation, however, is insufficient for the small eigenvalue end of the spectrum, where sub-leading terms play a significant role.  We extend the saddle point method to account for the sub-leading contributions.  We also develop a new approach, where the eigenvalue distribution is found as an equilibrium distribution at the endpoint of a stochastic process (Dyson Brownian motion).  The results of the two approaches are consistent in cases where both methods are applicable.  We discuss the implications of our results for vacuum stability and slow-roll inflation in the landscape.
}

\begin{document}
\maketitle
\flushbottom

\section{Introduction} 
\label{sect:Intro}

One of the striking predictions of string theory is the existence of a vast energy landscape with a multitude of vacuum states~\cite{Susskind:2003kw,BoussoPolchinski}.  
The landscape can be described by a multi-dimensional scalar potential $U({\bf \phi})$; then the vacua correspond to local minima of this potential.  In the cosmological context, positive-energy vacua drive the inflationary expansion of the universe, and transitions between different vacua occur by quantum tunneling through bubble nucleation.  The same kind of scenario is suggested by other particle physics models with compact extra dimensions.  For a review of this multiverse picture see, e.g., Ref.~\cite{LindeMultiverse}.  The expected number of vacua in the landscape is enormous, so predictions in this kind of theory must necessarily be statistical.  

The details of the high-energy vacuum landscape are not well understood, and it is often modeled as a random Gaussian field.  The statistics of vacuum energy densities and of slow-roll inflation in such a landscape have been extensively studied in the literature~\cite{Tegmark,Easther,Frazer,Battefeld,Bachlechner:2012at,Yang,Bachlechner,Wang,MV,EastherGuthMasoumi,MVY1,MVY2,MVY3,Bjorkmo:2017nzd,Blanco-Pillado:2017nin}.  Another well studied model is the axionic landscape, which can also be approximated by a random Gaussian field in a certain limit~\cite{
Kim:2004rp, Dimopoulos:2005ac, McAllister:2008hb, 
Higaki:2014pja, 
Wang:2015rel, Bachlechner:2017hsj}. 
One of the key mathematical problems to be addressed in these models is to find the eigenvalue distribution of the Hessian matrix $\zeta_{ij}=\partial^2 U/\partial\phi_i \partial\phi_j$.  The potential minima correspond to the points where $\partial U/\partial \phi_i =0$ and all Hessian eigenvalues are positive, and the stability of the vacuum depends on small eigenvalue end of the Hessian spectrum.  The dynamics of slow-roll inflation also depends on the smallest eigenvalues, which determine whether or not inflation is multi-field, with more than one field being dynamically important.     

The Hessian eigenvalue distribution in a random Gaussian field has been found in Ref.~\cite{Bray:2007tf} using the saddle point approximation in the leading order in the large-$N$ expansion, where $N$ is the dimensionality of the landscape.  This approximation, however, becomes inaccurate for very small eigenvalues, where sub-leading terms play a significant role.  In the present paper we extend the method of Ref.~\cite{Bray:2007tf} to account for the sub-leading contributions.\footnote{Alternative ways of going beyond the standard saddle point approximation have been discussed, in a different context, in Refs.~\cite{FyodorovNadal,FyodorovWilliams}. 
}  We also develop a new method, based on a version of Dyson Browninan motion \cite{Dyson:1962}, which can be applied in cases where the other method fails.  In this new approach, the eigenvalue distribution is obtained as an equilibrium distribution 
of the Brownian stochastic process. The results of the two approaches agree in cases where both methods are applicable.  We use our results to estimate the typical magnitude of the smallest Hessian eigenvalue at a local minimum of the potential and discuss its implications for the vacuum stability and for the dynamics of slow-roll inflation.  We also calculate the density of minima in a random Gaussian landscape. The result is consistent with earlier numerical calculations for $N\lesssim 100$ and extends them to larger values of $N$.

The paper is organized as follows. In the next section we specify the model of a random Gaussian landscape, review the probability distribution of the Hessian in this model, 
and clarify its relation to Wigner's random matrix model. 
In Sec.~\ref{sec:analytic}, we use the saddle point approximation to 
calculate the Hessian eigenvalue distribution at a generic point in the landscape, under the condition that all eigenvalues are larger than a given threshold. 
In Sec.~\ref{sec:analytic2}, we extend the analysis to stationary points of the landscape. We find the probability for a stationary point to be a minimum and estimate the smallest Hessian eigenvalue at a minimum. 
Then in Sec.~\ref{sec:dynamical} we develop a new method, based on Dyson Brownian motion, 
and use it to find the eigenvalue distribution at stationary points. 
Some cosmological implications of our results are discussed in Sec.~\ref{sec:implication}.
Our conclusions are summarized in Sec.~\ref{sec:conclusion}.  In Appendix~\ref{sec:axion} we discuss axionic landscapes and show that under certain conditions they can be approximated by random Gaussian fields.  We use the reduced Planck units ($M_{\rm pl} \simeq 2.4 \times 10^{18} \ {\rm GeV} \equiv 1$) throughout the paper.

\section{Random Gaussian Fields}
\label{sec:distribution}

\subsection{Correlators}

We consider a random Gaussian landscape $\pot (\boldsymbol \field)$, defined in an $N$-dimensional field space $\boldsymbol \field$, which is characterized by the average value ${\bar U}\equiv\langle \pot (\boldsymbol \field) \rangle$ and the correlation function 
\bel{Correlation}
\langle \pot (\boldsymbol \field_1) \pot(\boldsymbol \field_2)\rangle 
	- \bar{U}^2 
=  F (|\boldsymbol \field_1 - \boldsymbol \field_2|)=\frac1{(2\pi)^N}\int d^N {\bm k}\,P(k) e^{i{\bf k}\cdot (\boldsymbol\field_1-\boldsymbol\field_2)}~. 
\ee
Here, $k \equiv |\boldsymbol k|$ and angular brackets indicate ensemble averages.
We assume that the correlation function rapidly decays at $|\boldsymbol \field_1 - \boldsymbol \field_2| \gg \Lambda$ and the potential has a characteristic scale $U_0$. 
We define different moments of the spectral function $P(k)$ as 
\bel{sigmaDef}
	\sigma_{n}^2= \frac1{(2\pi)^N}\int d^N {\bm k} k^{2n} P(k)~. 
\ee
In Appendix~\ref{sec:axion}, we show that under certain conditions this type of random fields can be used to approximate axionic landscapes.

As an illustration, 
we may use the following correlation function: 
\beq
 F(\phi)=U_0^2 e^{-\phi^2/2\Lambda^2}, 
 \label{Gaussian F}
\eeq
with $\Lambda$ playing the role of the correlation length in the landscape. 
In this case, the moments are given by 
\beq
\sigma_n^2 = \frac{2^n \Gamma \lmk n + \frac{N}{2} \rmk }{\Gamma \lmk \frac{N}{2} \rmk} 
\frac{U_0^2}{\Lambda^{2n}}. 
\eeq
In the large-$N$ limit the moments are of the order 
\be
\sigma_n^2 \sim U_0^2 \left({N}\over{\Lambda^2}\right)^n. 
\label{sigma}
\ee
In the rest of this paper, we do not use the above explicit form of the correlation function, 
but generically assume only the dependence of \eq{sigma}.

Let us consider the potential around a given point in the field space and expand it in  a Taylor series.  Since the values of the potential at nearby points are correlated with one another, the coefficients of the Taylor expansion should also be correlated.  In particular we have 
\beq
 && \la U ({\bm \phi} ) \ra \equiv \bar{U} 
\label{1}
 \\
 && \la (U ({\bm \phi} )  - \bar{U})^2 \ra
 = E 
 \\
 &&\la U ({\bm \phi}) \zeta_{ij} ({\bm \phi}) \ra
 = 
 B \delta_{ij} 
 \\
 &&\la \zeta_{ij} ({\bm \phi}) \zeta_{kl} ({\bm \phi}) \ra
 = 
 A \lmk \delta_{ij} \delta_{kl} + \delta_{ik} \delta_{jl} + \delta_{il} \delta_{jk} \rmk
 \label{correlation of Hessian}
 \\
 && \la \eta_i ({\bm \phi}) U({\bm \phi})\ra =  \la \eta_i ({\bm \phi}) \zeta_{ij}({\bm \phi})\ra =0,
 \label{2}
 \eeq
where $\eta_i= \del U/\del \phi_i$ and $\zeta_{ij} \equiv \del^2 U / \del \phi_i \del \phi_j$ is the Hessian matrix. The parameters $E, B, A$ are related to the moments (\ref{sigmaDef}) as 
\beal{varDef}
	E=\sigma_0^2~, 	~~~~
	A= \frac{\sigma_2^2}{N(N+2)}~, ~~~~
	B= -\frac1N \sigma_1^2~. 
\eea
From \eq{sigma}, we expect that $A, B, E$ are $\mathcal{O}(N^0)$ in the large $N$ limit.

\subsection{Probability distribution}

The probability distribution for $U$ and $\zeta_{ij}$ can be found by taking 
the inverse of the correlation matrix. The resulting distribution is~\cite{Bray:2007tf},\cite{MVY1}
\bel{jointUHess}
	P(\pot, \hs)\propto e^{-Q_{\pot, \hs}}~,
\ee
where
\bea
	&&Q_{\pot, \hs} = \frac{(N+2)A}{(N+2) AE - NB^2} \lmk 
	\frac{1}{2} \lmk \pot - \bar{U} \rmk^2 -  \frac{B}{(N+2)A}  \lmk \pot - \bar{U} \rmk \Tr \hs 
	\right.
	\nn
	&&~~~~~~~~~~~~~~~~~~~~~~~~~~~~~~~~~~~~~~~~~~~~~~~~~~\left.
	- \frac{A E - B^2}{4(N+2)A^2} (\Tr \hs)^2 \rmk + \frac1{4A} \Tr \hs^2~. 
\label{QUzeta}
\eea
Note that the combination $AE - B^2$ must be positive (or zero), since otherwise the distribution cannot be normalized.  

The cross term in Eq.~(\ref{QUzeta}) can be absorbed by a constant shift of the eigenvalues of the Hessian using the relations 
\beq
	\lkk \Tr \lmk \hs - \lambda_* I \rmk \rkk^2 =
	\lmk \Tr \hs \rmk^2 - 2 N \lambda_* \Tr \hs + N^2 \lambda_*^2 
	\\
	\Tr \lkk \lmk \hs - \lambda_* I \rmk^2 \rkk= \Tr \lmk \hs^2 \rmk - 2 \lambda_* \Tr \hs + N \lambda_*^2, 
\eeq
and setting 
\beq
 \lambda_* (U) = \frac{B}{E} \lmk U - \bar{U} \rmk, 
 \label{lambda_*}
\eeq
where $I$ is the identity matrix. 
This implies that the eigenvalues of the Hessian are shifted by amount of the order $- U/ \Lambda^2$ for a given $U$, where we have used $B < 0$. 

The Hessian matrix can be diagonalized and the distribution can be written in terms of its eigenvalues $\lambda_i$. 
Changing the variables from $\zeta_{ij}$ to $\lambda_i$, 
the probability distribution for the eigenvalues is given by~\cite{Dyson:1962}
\beq
 P(\lambda) = C \prod_{i < j} \abs{\lambda_i - \lambda_j} 
 e^{ - Q_{U,\zeta}}
 \label{probability of lambda}
\eeq
where $\prod \abs{\lambda_i - \lambda_j}$ comes 
from the Jacobian and $C$ is a normalization factor. 

We denote the average eigenvalue as $\bar{\lambda}$ 
and deviations from the average as $\delta \lambda_i$: 
\beq
 &&\lambda_i = \bar{\lambda} + \delta \lambda_i 
 \\
 &&\sum_i \delta \lambda_i = 0. 
\eeq
Then $Q_{U, \hs}$ can be written as 
\beq
 Q_{U, \hs} \simeq 
 \frac{1}{4 A} \sum_i \delta \lambda_i^2 
 + \frac{E}{2 (AE - B^2) } \lmk \bar{\lambda} - \lambda_* (U) \rmk^2 
 + {\rm const} 
\label{Q final}
\eeq
for a fixed $U$, where we have used 
\beq
 - \frac{1}{(N+2) AE - NB^2} \frac{A E - B^2}{4A} 
 \simeq - \frac{1}{4 A N} + \frac{E}{2 N^2 (AE - B^2) } 
 + \mathcal{O}( N^{-3}) \times \frac{1}{A}, 
\eeq
in the large $N$ limit.%
\footnote{
This expansion is not a good approximation for the Gaussian correlation function 
(\ref{Gaussian F}) 
because $AE - B^2 = 0$ in that case. 
We do not focus on this particular case 
but consider a generic situation where $AE - B^2 = {\cal O}(1)$. 
However, \eq{Q final2} is still correct for any correlation functions. 
}
Note that the Jacobian that appears in \eq{probability of lambda} 
is independent of $\bar{\lambda}$.

One may be interested in the probability distribution for the Hessian eigenvalues without any condition on $U$. In this case, $U$ can be integrated out and the distribution takes the form \cite{Fyodorov}
\bea
	Q_{\hs} = 
 \frac1{4A} 
 \lkk \Tr \hs^2
 - \frac{1}{N+2} (\Tr \hs)^2 
 \rkk ,
 \label{Q final2}
\eea
or 
\beq
 Q_{\hs} \simeq 
 \frac{1}{4 A} 
 \lmk 
 \sum_i \delta \lambda_i^2 
 + \frac{2N}{N+2} \bar{\lambda}^2 
 \rmk. 
\label{Q final3}
\eeq

Here we comment on the difference from the random matrix theory (RMT), 
where the probability distribution for the elements of a real symmetric matrix $\zeta_{ij}$
is given by \cite{Wigner}
\beq
 P_\hs = \frac{1}{\lmk \sqrt{2 \pi} \sigma_{\rm RMT} \rmk^{N(N+1)/2}} e^{-Q_{\rm RMT}} 
 \\
 Q_{\rm RMT} = \frac{1}{2 \sigma_{\rm RMT}^2} \Tr \hs^2, 
\eeq
with a certain constant $\sigma_{\rm RMT}$.  This distribution is usually referred to as the Gaussian Orthogonal Ensemble (GOE).
When we express the eigenvalues of $\zeta$ in terms of the average value $\bar{\lambda}$ 
and displacements from the average $\delta \lambda_i$, 
the exponent $Q_{\rm RMT}$ is rewritten as 
\beq
 Q_{\rm RMT} = \frac{1}{2 \sigma_{\rm RMT}^2} 
 \lmk \sum_i \delta \lambda_i^2 
 + N \bar{\lambda}^2 
 \rmk. 
 \label{Q_RMT decomposed}
\eeq
To compare $Q_{U, \hs}$ (or $Q_{\hs}$) and $Q_{\rm RMT}$ we may take $\sigma_{\rm RMT}^2 = 2 A$ ($\sim U /\Lambda^2$). 
We then see that the cost of a nonzero $\bar{\lambda}$ in the GOE is larger than that for a random Gaussian field by a factor of $N$. 
Thus, for the GOE, the averaged value $\bar{\lambda}$ is strongly prohibited 
from being away from zero in the large-$N$ limit.

Let us emphasize that the Gaussian correlation function (\ref{Gaussian F}) is a rather special example of a random Gaussian model. 
It has a specific property because the coefficient of $(\Tr \zeta)^2$ in (\ref{QUzeta}) vanishes ($AE - B^2 = 0$).  As a result, for a fixed $U$ 
the Hessian distribution is just given by the GOE with a constant shift 
of the diagonal terms:  
\beq
 \zeta_{ij} = m_{ij} -  \frac{B}{E} (U - \bar{U}) \delta_{ij}, 
 \label{shifted GOE}
\eeq
where $m_{ij}$ is a GOE matrix. 
However, this is not a generic property of random Gaussian models. 
In what follows, we do not consider this special case, but consider a generic random Gaussian landscape, which is specified by moments of the correlation function.

\section{Saddle point approximaion}
\label{sec:analytic}

In this section we use the saddle point approximation to calculate the probability distribution of Hessian eigenvalues in a random Gaussian landscape under the condition that all eigenvalues are greater than a given threshold.  We follow and extend the calculation of Refs.~\cite{Dean:2006wk, DeanMajumdar}, 
where the eigenvalue distribution was found for the case of the GOE. 
In Sec.~\ref{sec:analytic1}, we calculate the distribution at a generic point in the landscape. 
The distribution at local minima of the potential cannot be found with this method.  However, 
the result of this calculation can be used to find the probability for a stationary point of the potential to be a minimum and to estimate the smallest Hessian eigenvalue at a minimum, 
as we will see 
in Sec.~\ref{sec:minima}. The calculation of the Hessian eigenvalue distribution at potential minima should await the introduction of our new method in Sec.~\ref{sec:dynamical}.

Hereafter, we generically consider the case where 
\beq
 Q_\zeta = \frac{1}{2} \lkk \Tr \zeta^2 - \frac{a}{N} \lmk \Tr \zeta \rmk^2 \rkk, 
 \label{Q_zeta}
\eeq
which can be obtained from \eq{Q final2} by rescaling 
$\lambda_i \to \sqrt{2A} \lambda_i$ with $a =N/(N+2)$. 
It can also represent (\ref{Q final}) with $a = 1 - 2AE / N (AE - B^2)$ and the same rescaling, 
after a shift $\lambda_i \to \lambda_i + \lambda_*$.  The latter case will be discussed in detail in Sec.~\ref{sec:fixedU}.
In both cases, we expect $1 - a = \mathcal{O}(N^{-1})$. 

\subsection{Conditional probability distribution}
\label{sec:analytic1}

The probability distribution for the Hessian eigenvalues can be written as
\beq
  &&p({\bm \lambda}) = A \exp \lmk - H ({\bm \lambda}) \rmk 
 \label{Boltz}
  \\  
  &&H({\bm \lambda}) =  \frac{1}{2} \lmk \sum_i \lambda_i^2 
  - \frac{a}{N} \lkk \sum_i \lambda_i \rkk^2  - \sum_{i \ne j} \ln \lmk \abs{\lambda_i - \lambda_j} \rmk 
  \rmk, 
  \label{Hamiltonian}
\eeq
where $A$ is a normalization constant and the logarithmic term comes from the Jacobian factor in Eq.~(\ref{probability of lambda}). 
The conditional probability $P( \lambda_{\rm cr})$ that all eigenvalues are greater than some value $\lambda_{\rm cr}$ can then be calculated from 
\beq
 &&P (\lambda_{\rm cr}) = \frac{Z (\lambda_{\rm cr})}{ Z( - \infty)} 
 \label{Q}
 \\
 &&Z (\lambda_{\rm cr}) = \int_{\lambda_{\rm cr}}^\infty \dd^N {\bm \lambda}
 \exp 
 \lmk - H ({\bm \lambda}) \rmk. 
 \label{def:Z_cr}
\eeq

We shall further rescale the eigenvalues as ${\bm \mu} = {\bm \lambda} / \sqrt{N}$ 
and introduce a density function of ${\bm \mu}$ 
as 
\beq
 \rho( \mu) = \frac{1}{N} \sum_i \delta \lmk \mu - \mu_i \rmk. 
\eeq
In terms of this density function, we can rewrite $H({\bm \lambda})$ as 
\beq
 H [ \rho] / N^2 &=& \frac{1}{2} \int \dd \mu \mu^2 \rho (\mu) 
 - \frac{1}{2} a \int \dd \mu \dd \mu' \rho(\mu) \rho(\mu') \mu \mu'
 \nonumber\\
 &&- \frac{1}{2} \int \dd \mu \dd \mu' \rho(\mu) \rho(\mu') \ln \lmk \abs{\mu - \mu'} \rmk. 
\eeq

The partition function $Z (\lambda_{\rm cr})$ can also be rewritten in terms of $\rho (\mu)$. 
The Jacobian involved in changing from $\mu_i$ to $\rho (\mu)$ 
was calculated in Ref.~\cite{Dean:2006wk, DeanMajumdar} by saddle point approximation 
in the large $N$ limit. It is given by 
\beq
 J [\rho] 
 &=& A' \int \prod_{i=1} \dd \mu_i \delta \lkk N \rho(\mu) - 
 \sum_i \delta (\mu - \mu_i) \rkk 
 \\
 &\simeq& A'' \delta \lmk \int \dd \mu \rho (\mu) - 1 \rmk \exp \lkk - N \int \dd \mu \rho (\mu) \ln \rho \rkk, 
 \label{Jacobian2}
\eeq
where $A'$ and $A''$ are normalization constants. 
Thus we obtain 
\beq
 Z (\lambda_{\rm cr}) &=& 
 A''' 
 \int \dd C \dd [\rho] 
e^{- N^2 \Sigma [ \rho]  }
 \label{Z_cr}
\\
 \Sigma_0 [\rho] &=& 
 \frac{1}{2} \int \dd \mu \mu^2 \rho (\mu) 
 - \frac{1}{2} \int \dd \mu \dd \mu' \rho(\mu) \rho(\mu') \mu \mu'
 \nonumber\\
 &&- \frac{1}{2} \int \dd \mu \dd \mu' \rho(\mu) \rho(\mu') \ln \lmk \abs{\mu - \mu'} \rmk 
 + C \lkk \int \dd \mu \rho (\mu) - 1 \rkk 
 \label{Z_0}
 \\
 \Sigma_1 [ \rho] &=& 
   \frac{1}{2} N(1-a) \int \dd \mu \dd \mu' \rho(\mu) \rho(\mu') \mu \mu' 
 + \int \dd \mu \rho (\mu) \ln \lkk \rho (\mu) \rkk, 
\label{Sigma1}
\eeq
where 
\beq
\Sigma [\rho] = \Sigma_0 [\rho] + \Sigma_1[\rho] /N + \mathcal{O}(1/N^2),
\label{sigmas}
\eeq 
$A'''$ is a normalization constant, and we include a Lagrange multiplier $C$ to set the normalization of $\rho (\mu)$ coming from the delta function in \eq{Jacobian2}. 
Note that $N (1-a) = {\cal O}(1)$. 
The functional integration in (\ref{Z_cr}) is over functions $\rho(\mu)$ satisfying $\rho(\mu)=0$ for $\mu<\mu_{\rm cr}$, where $\mu_{\rm cr}=\lambda_{\rm cr}/\sqrt{N}$.

\subsubsection{Eigenvalue density function}

We shall now use the saddle point approximation to find the most probable density function $\rho(\mu)$. We first note that the leading term $\Sigma_0[\rho]$ in Eq.~(\ref{sigmas}) is independent of ${\bar\lambda}$.  We therefore include the subleading contribution due to the first term in $\Sigma_1[\rho]$, which breaks this degeneracy.  The second term in $\Sigma_1[\rho]$ is also independent of ${\bar\lambda}$, and we shall neglect it here.  This term only gives 
${\cal O}(N^{-7/4})$ corrections to $\Delta \Sigma$ ($\equiv \Sigma (\mu_{\rm cr}) - \Sigma(-\infty)$), as we will discuss later in this section.

Varying the functional $\Sigma[\rho]$ with respect to $\rho (\mu)$, we determine 
the critical distribution $\rho_c (\mu)$ at the saddle point, 
\beq
 \frac{\mu^2}{2} - a \mu \int_{\mu_{\rm cr}}^\infty  \dd \mu' \rho_{\rm c}(\mu') \mu' + C = 
 \int_{\mu_{\rm cr}}^\infty \dd \mu' \rho_{\rm c} (\mu') \ln \lmk \abs{\mu - \mu'} \rmk. 
 \label{saddle point}
\eeq
Taking a derivative with respect to $\mu$, we obtain 
\beq
 \mu 
 - a \int_{\mu_{\rm cr}}^\infty  \dd \mu' \rho_{\rm c}(\mu') \mu' 
 = \mathcal{P} \int_{\mu_{\rm cr}}^\infty \dd \mu' \frac{\rho_{\rm c} (\mu')}{\mu - \mu'}, 
 \label{saddle point2}
\eeq
where $\mathcal{P}$ indicates the Cauchy principal part. 
Shifting $\mu$ as $\mu = x + \mu_{\rm cr}$, this can be rewritten as 
\beq
 x + x_0 = 
 \mathcal{P} \int_0^\infty \dd x' \frac{\rho_{\rm c} (x' + \mu_{\rm cr})}{x - x'}, 
\label{integraleq}
\eeq
where we have defined 
\beq
 x_0 (\mu_{\rm cr}) \equiv 
 \mu_{\rm cr} - a \int_0^\infty  \dd x' \rho_{\rm c}(x' + \mu_{\rm cr}) (x' + \mu_{\rm cr}). 
 \label{x_0}
\eeq
The integral in the last equation is just the average eigenvalue ${\bar\mu}$; hence this equation can also be written as
\beq
x_0 =  \mu_{\rm cr} - a{\bar\mu}.
\label{x0barmu}
\eeq

The solution of the integral equation (\ref{integraleq}) has been found in Ref.~\cite{Dean:2006wk, DeanMajumdar}. Here we quote the result: 
\beq
 \rho_{\rm c} ( x + \mu_{\rm cr}) = 
 \frac{1}{2 \pi \sqrt{x}} \sqrt{L( x_0) - x} \lkk L(x_0) + 2x + 2x_0 \rkk. 
 \label{rho_c}
\eeq
This solution applies in the range $x \in [ 0, L (x_0)]$. Otherwise $\rho_{\rm c} ( x + \mu_{\rm cr}) = 0$.  The function $L(x_0)$ is determined by the normalization 
\beq
 \int_0^L \rho_{\rm c} \dd x = 1. 
 \label{normalization}
\eeq
Then we obtain 
\beq
 L(x_0) = \frac{2}{3} \lkk \sqrt{x_0^2 + 6} - x_0 \rkk. 
\eeq

In the special case when $\mu_{\rm cr}=-\sqrt{2}$, we have $x_0=-\sqrt{2}$, $L(x_0)=2\sqrt{2}$ and
\beq
 \rho_W ( x + \mu_{\rm cr})= \pi^{-1}\left(2\sqrt{2} x-x^2\right)^{1/2}
 \eeq
 or
 \beq
 \rho_W(\mu)=\pi^{-1}\left(2-\mu^2\right)^{1/2}.
 \label{Wigner}
 \eeq 
This is the celebrated Wigner semi-circle distribution.  It has support at $-\sqrt{2}<\mu <\sqrt{2}$, and thus the requirement $\mu>\mu_{\rm cr}$ with $\mu_{\rm cr}=-\sqrt{2}$ does not impose any constraint on $\rho(\mu)$.  For the same reason the Wigner distribution is unperturbed when $\mu_{\rm cr}<-\sqrt{2}$.

\begin{figure}[t] %  figure placement: here, top, bottom, or page
   \centering
   \includegraphics[width=4.5in]{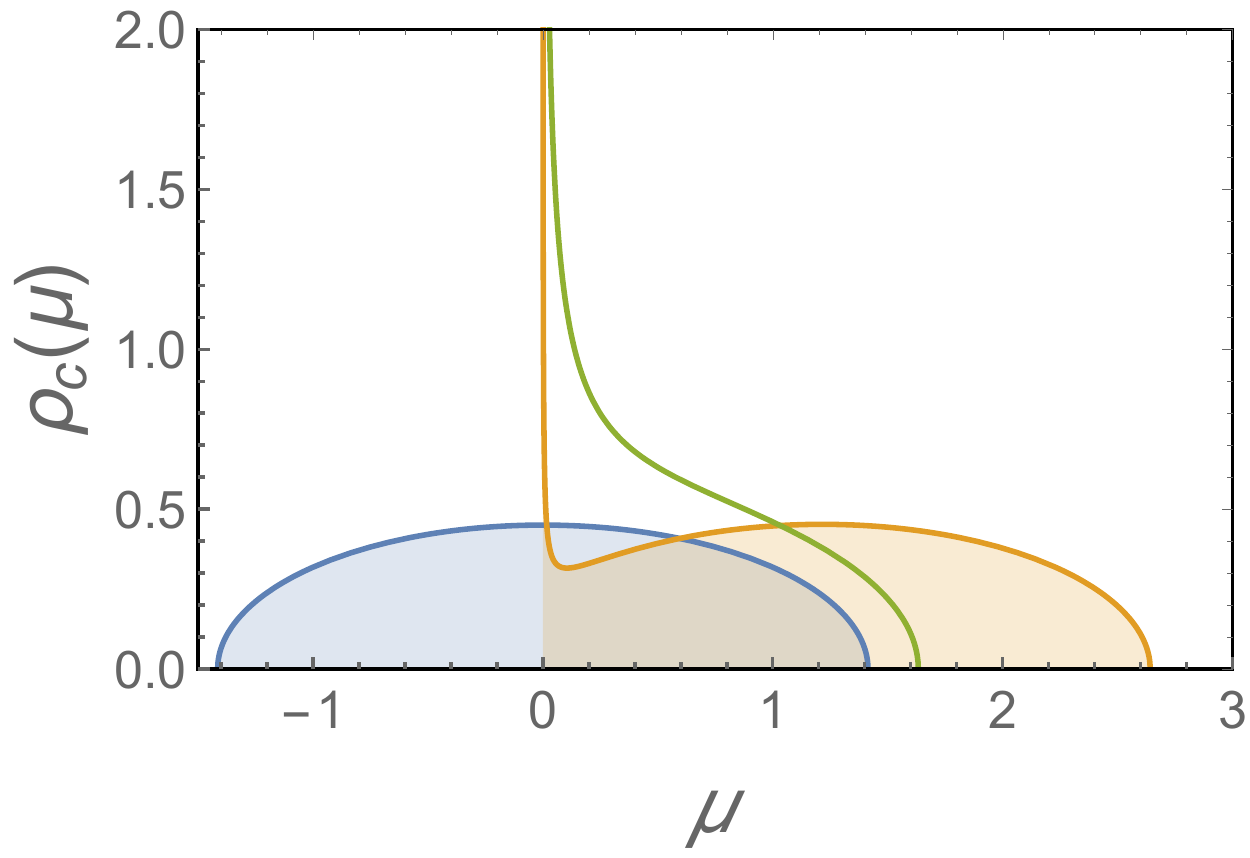} 
   \caption{
Eigenvalue distribution $\rho(\mu)$ restricted to $\mu >0$ in the random matrix theory (green curve) and for the Hessian in a random Gaussian landscape (orange curve) with $N=100$.
The blue curve is the distribution without any restrictions, which is given by the Wigner semi-circle.  
   }
   \label{fig:semi0}
\end{figure}

The Gaussian orthogonal ensemble (GOE), which was studied in Refs.~\cite{Dean:2006wk, DeanMajumdar}, corresponds to $a=0$; then Eq.~(\ref{x_0-2}) gives $x_0=0$ for $\mu_{\rm cr} = 0$.  The eigenvalue distribution for this case is shown by a green curve in Fig.~\ref{fig:semi0}.

\begin{figure}[t] %  figure placement: here, top, bottom, or page
   \centering
   \includegraphics[width=2.5in]{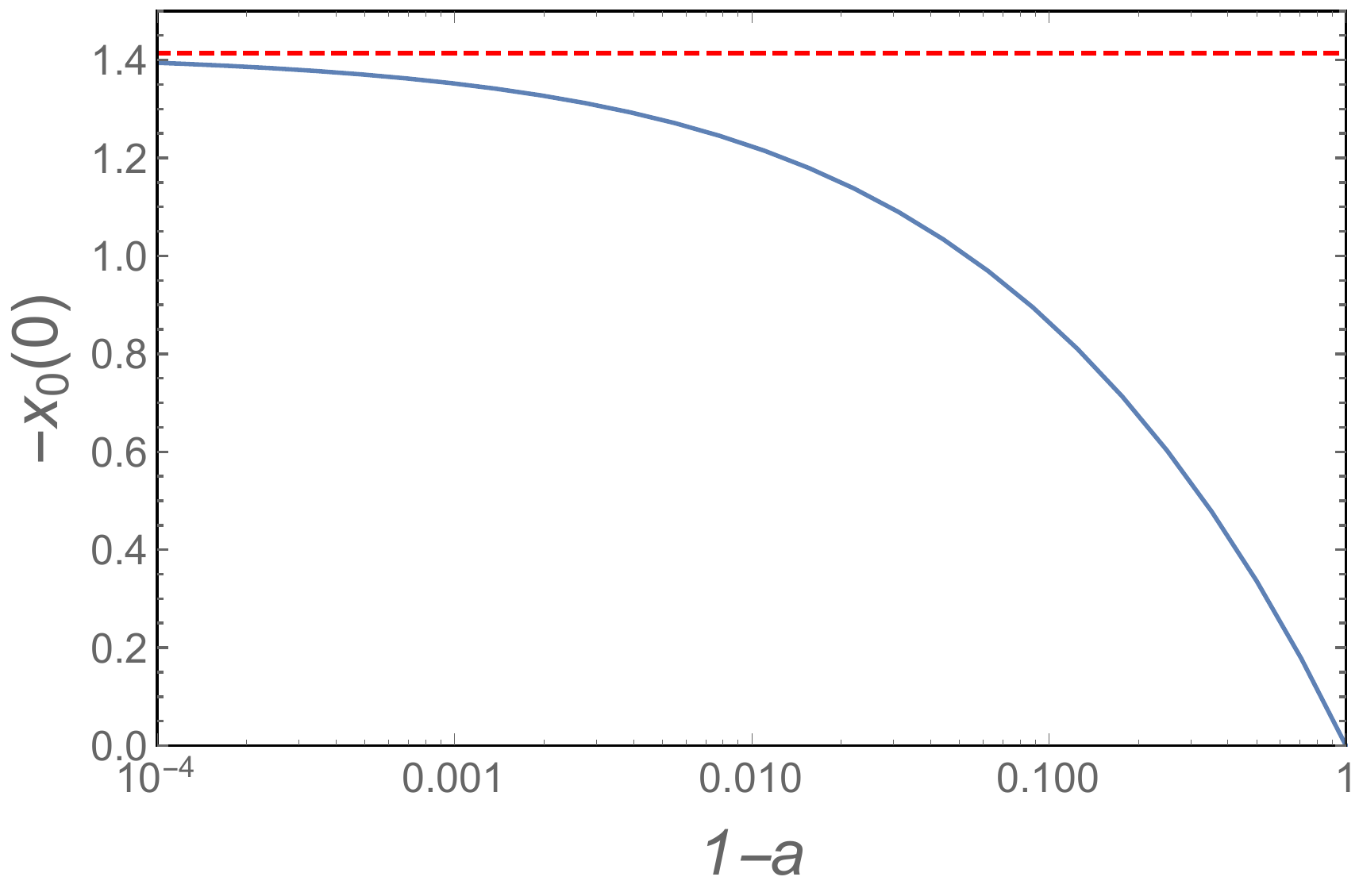} 
   \caption{
$-x_0(\mu_{\rm cr} = 0)$ as a function of $(1-a)$.
The red dashed line represents the asymptotic value $-x_0 = \sqrt{2}$ 
in the limit of $1-a \to 0$. 
   }
   \label{fig:x0}
\end{figure}

The integral in \eq{x_0} can be done by using the explicit forms of $\rho_{\rm c}$ and $L(x_0)$. 
As a result we obtain
\beq
 F(x_0) = (1-a) \lmk \mu_{\rm cr} + F(x_0) - x_0 \rmk, 
 \nonumber \\\label{x_0-2}
\eeq
where
\beq
F(x) \equiv \frac{1}{27} \left[-x(x^2+9)+\left(6+x^2\right)^{3/2}\right].
\label{Fx}
\eeq
We can numerically solve \eq{x_0-2} in terms of $x_0$ for given values of $a$ and $\mu_{\rm cr}$.  One is often interested in the case when $\mu_{\rm cr} = 0$, so that all eigenvalues are positive.  In Fig.~\ref{fig:x0} we show $x_0$ as a function of $(1-a)$ for $\mu_{\rm cr} =0$. 
We see that $-x_0$ asymptotes to $\sqrt{2}$ (as indicated by the red dashed line) for $1-a \to 0$.  
To clarify the asymptotic behavior, we Taylor expand the function $F(x)$ about $x=-\sqrt{2}$.  This gives
\beq
F(x)=\frac{1}{2\sqrt{2}}\left(x+\sqrt{2}\right)^2 + \dots. 
\eeq
Now, to the leading order in $(1-a)$, Eq.~(\ref{x_0-2}) becomes
\beq
\frac{1}{2\sqrt{2}}\left(x_0+\sqrt{2}\right)^2=(1-a) (\mu_{\rm cr} + \sqrt{2}),
\eeq
where we have used that $F(-\sqrt{2})=0$.  Hence we find
\beq
x_0 = -\sqrt{2}+2^{3/4} (1-a)^{1/2} \left(\mu_{\rm cr}+\sqrt{2}\right)^{1/2}.
\label{x_0 approximation1}
\eeq
For $\mu_{\rm cr}=0$ this gives
\beq
x_0 = -\sqrt{2}+2\sqrt{1-a}.
\label{x_0 approximation}
\eeq

It is interesting to note that $x_0 = -\sqrt{2}$ for $1-a = 0$ and any value of $\mu_{\rm cr}$. 
In this case Eq.~(\ref{x0barmu}) gives ${\bar\mu} =\sqrt{2}+\mu_{\rm cr}$.
This means that the distribution $\rho_c (\mu)$ is just given by a shifted Wigner semi-circle (\ref{Wigner}) 
when we neglect the next-leading order correction $\Sigma_1 [\rho]$ in the large $N$ limit \cite{{Bray:2007tf}}. Deviations from the Wigner semi-circle come from the next-leading order effect. 

\begin{figure}[t] %  figure placement: here, top, bottom, or page
   \centering
   \includegraphics[width=2.5in]{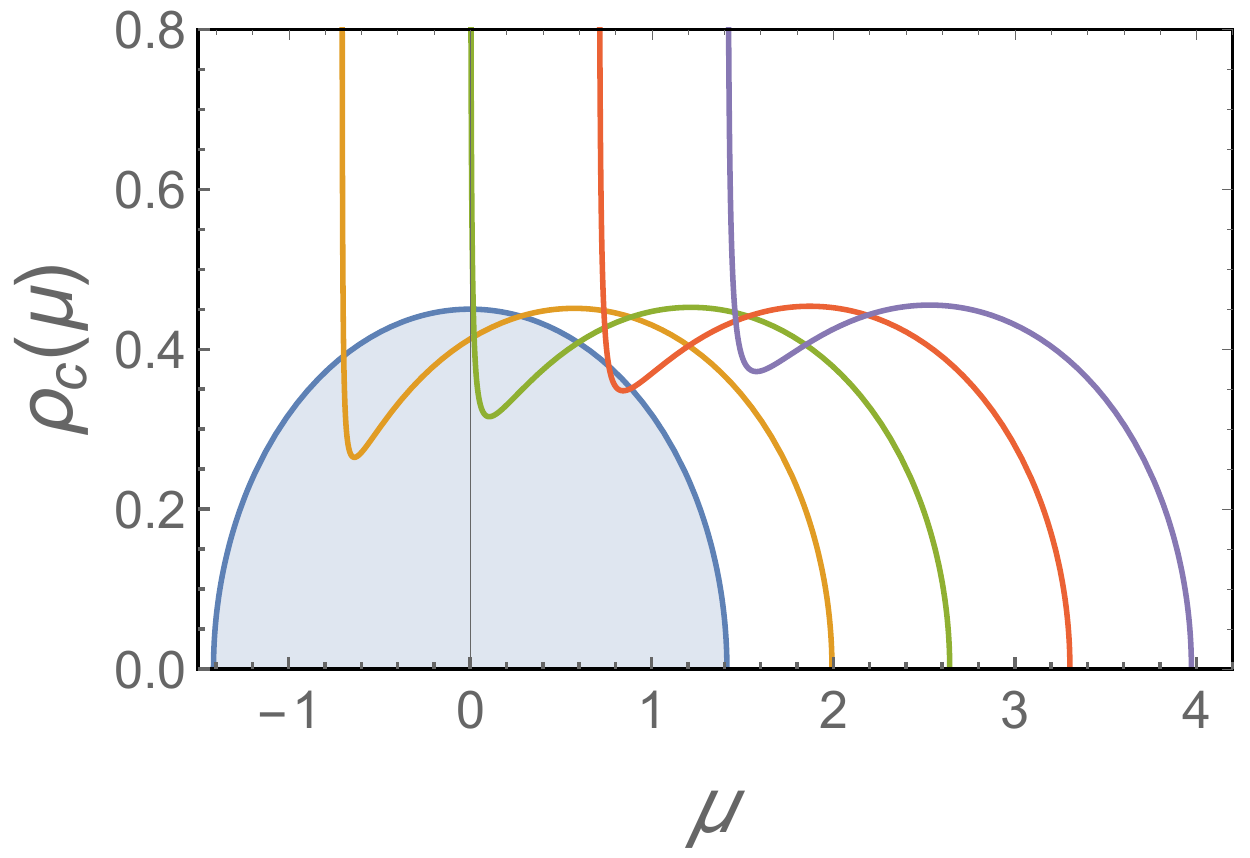} 
   \quad 
   \includegraphics[width=2.5in]{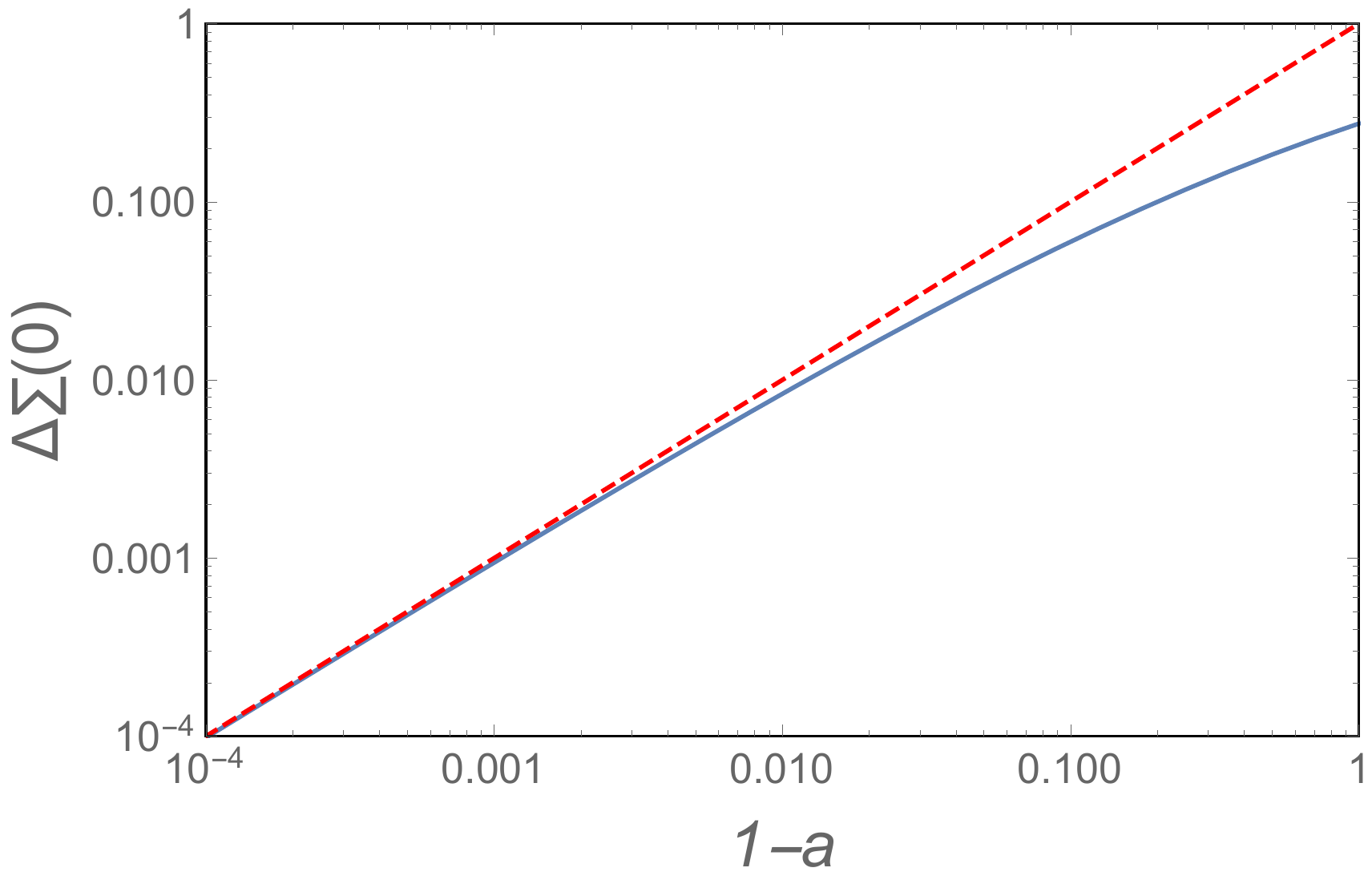} 
   \caption{
   {\bf Left:} $\rho_c (\mu)$ as a function of $\mu$ with $a = N/(N+2)$ and $N=100$, 
   where we take $\mu_{\rm cr} = -\sqrt{2}, \, -\sqrt{2}/2, \, 0 , \, \sqrt{2}/2, \, \sqrt{2}$ 
   from left to right. The case of $\mu_{\rm cr} = -\sqrt{2}$ is filled in blue color 
   and is given by the Wigner semi-circle. 
   {\bf Right:} $\Delta \Sigma [ \rho]$ as a function of $1-a$ with $\mu_{\rm cr}  = 0$.
   The asymptotic form is shown by a dashed red line. 
   }
   \label{fig:sigma}
\end{figure}

In the left panel of Fig.~\ref{fig:sigma}, we plot the eigenvalue distribution $\rho_c (\mu)$ with $\mu_{\rm cr} = - \sqrt{2}, -\sqrt{2}/2, 0, \sqrt{2}/2, \sqrt{2}$ for the case of $a = N/(N+2)$ (which corresponds to the Hessian distribution (\ref{Q final2}) and $N=100$. 
We also plot the distribution for $\mu_{\rm cr} =0$ as an orange curve in Fig.~\ref{fig:semi0}, to compare it with the GOE distribution (plotted as a green curve).  We see that the GOE distribution is much more concentrated near the origin, reflecting the high cost of a nonzero average eigenvalue ${\bar\mu}$ in that case.

The Hessian distributions in Fig.~\ref{fig:sigma} look like the Wigner semi-circle with an overall shift and a slight modification at the left edge. 
From Eqs.~(\ref{x0barmu}) and (\ref{x_0 approximation}), the amount of the shift can be estimated as 
\beq
\bar{\mu} \approx \sqrt{2} - 2 \sqrt{1-a}
\eeq 
for $\mu_{\rm cr} = 0$ and 
in the limit of $(1-a) \ll 1$.  The form of the distribution near the left edge, $0<\mu\lesssim 2\sqrt{1-a}$, where it significantly deviates from the Wigner semi-circle, can be found from 
Eqs.~(\ref{rho_c}) and (\ref{x_0 approximation}): 
\beq
 \rho_c (\mu) \simeq \frac{5}{3 \pi \sqrt{\mu}} \sqrt{ 2 \sqrt{2} (1-a)}. 
 \label{left edge}
\eeq
The number of eigenvalues in this range is
\beq
 N \int_0^{2 \sqrt{1-a}} \rho_c (\mu) \dd \mu \sim \frac{20 \cdot  2^{1/4}}{3 \pi}  N (1-a)^{3/4} \sim 
 N^{1/4}. 
 \label{N_epsilon2}
\eeq

\subsubsection{Probability of $\mu>\mu_{\rm cr}$}

The partition function $Z(\mu_{\rm cr})$ can be approximated by its value 
at the saddle point: 
\beq
 Z (\mu_{\rm cr}) \sim 
 \exp \lkk - N^2 \Sigma (\mu_{\rm cr}) \rkk. 
\eeq
Using \eq{saddle point}, 
$\Sigma (\mu_{\rm cr})$ is given by 
\beq
 \Sigma (\mu_{\rm cr}) 
 = \frac{1}{4} \int_0^{L(x_0)} \dd x (x + \mu_{\rm cr})^2 \rho_{\rm c} (x+\mu_{\rm cr}) 
 - \frac{1}{2} C. 
 \label{Sigma_0-3}
\eeq
Here, the Lagrange multiplier $C$ can be determined from \eq{saddle point} 
by setting $\mu = \mu_{\rm cr}$, 
\beq
 - \frac{1}{2} C &=& \frac{1}{4} \mu_{\rm cr}^2 
 - \frac{a}{2} \mu_{\rm cr} \int_0^{L(x_0)} \dd x (x + \mu_{\rm cr}) \rho_{\rm c} (x+\mu_{\rm cr}) 
 \nonumber\\
 &&- \frac{1}{2} \int_0^{L(x_0)} \dd x \ln x \rho_{\rm c} (x+\mu_{\rm cr}). 
\eeq
These integrals can be done explicitly by using \eq{rho_c}. 
The result is  
\beq
  &&\Sigma (\mu_{\rm cr})
 = \frac{1}{864} \lmk - x_0 + \sqrt{6 + x_0^2} \rmk^3 \lmk x_0 + 3 \sqrt{6  + x_0^2} \rmk 
 + \frac{1- a}{2a} \mu_{\rm cr} (\mu_{\rm cr} - x_0) 
 \nonumber\\
 &&~~~~
 + \frac{1}{12} 
 \lkk 3 + x_0 \lmk - x_0 + \sqrt{6 + x_0^2} \rmk + 3 \ln 36 - 6  \ln \lmk -x_0 + \sqrt{6  +x_0^2} \rmk 
 \rkk, 
 \label{Sigma_c}
\eeq
where we have used \eq{x_0}. 

Since we are interested in the case where $(1-a) \ll 1$, 
we can simplify \eq{Sigma_c} by using \eq{x_0 approximation1}. 
The result is  
\beq
  \Sigma (\mu_{\rm cr}) = \frac{1}{8} \lmk 3 + \ln 4 \rmk 
  + \frac{1}{2} \lmk \mu_{\rm cr} + \sqrt{2} \rmk^2 (1-a) 
  + {\cal O}( (1-a)^{3/2}). 
  \label{sigma-approximation}
\eeq

The probability for all eigenvalues to be greater than $\mu_{\rm cr}$ is given by $P_>(\mu_{\rm cr}) = \exp [ - N^2 \Delta \Sigma (\mu_{\rm cr})]$, 
where $\Delta \Sigma (\mu_{\rm cr}) \equiv \Sigma (\mu_{\rm cr}) - \Sigma (- \infty)$ 
and 
$\Sigma (- \infty)  = (3 + \ln 4)/8$. 
This probability can be found numerically using Eqs.~(\ref{x_0}) and (\ref{Sigma_c}).
In the right panel of Fig.~\ref{fig:sigma}, we plot $\Delta \Sigma (0)$ as a function of $(1-a)$. 
We also plot the asymptote $\Delta \Sigma (0) = (1-a)$ in the limit of $(1-a) \to 0$
as a dashed red line. 
In the case of the Hessian distribution (\ref{Q final2}), $(1-a) = 2/(N+2)$, we obtain $P_>(0) \sim e^{-2N^2/(N+2)}$.%
\footnote{
This can be derived from the result of Ref.~\cite{Bray:2007tf} if we take a
limit of $\alpha \to 0$, where $\alpha$ is defined as a fraction of
eigenvalues which are negative. However, their analysis is inaccurate in that limit. See also discussion below \eq{probmin}.}

Now we can justify that the second term of $\Sigma_1$ in Eq.~(\ref{Sigma1})
gives a negligible contribution to $\Delta \Sigma$ in the large $N$ limit.  As we already mentioned, this term is independent of the average eigenvalue ${\bar\mu}$.  
Furthermore, from \eq{N_epsilon2} we see that the change in $\Sigma_1$ due to the modified distribution near $\mu=0$
is of the order $N^{1/4} / N$. 
It follows that the contribution of the second term of $\Sigma_1$ to $\Delta \Sigma$ is 
$\mathcal{O}(N^{-7/4})$, which is much smaller than the other terms 
in the large $N$ limit. 
We checked that it is indeed $\mathcal{O}(N^{-7/4})$ by numerically calculating $N^{-1} \int d \mu \rho \ln \rho$ as a function of $N$ with $\rho$ given by \eq{rho_c}. 
Therefore, our result for $\Delta \Sigma$ is accurate with an uncertainty of $\mathcal{O}(N^{-7/4})$. 
Additional support for neglecting the second term of $\Sigma_1$ comes from the fact that 
the distribution (\ref{rho_c}) obtained without this term agrees 
very well with the result of the dynamical method, which takes all terms into account (see Sec.~\ref{sec:dynamical}).

\section{Probability of $\mu>\mu_{\rm cr}$ at stationary points of the potential}
\label{sec:analytic2}

We shall now calculate the probability for all Hessian eigenvalues to be greater than a given value $\mu_{\rm cr}$ at stationary points, where $\del_i U = 0$.  For $\mu_{\rm cr}=0$, this is the same as the probability for a stationary point to be a local minimum.
We insert a delta function in \eq{def:Z_cr} for the partition function to enforce the condition $\del_i U = 0$ in the landscape: 
\beq
 \int \prod_i \dd \phi_i 
 \delta ( \del_i U) \abs{\det \zeta}. 
 \label{deltafunction}
\eeq
The Jacobian $\abs{\det \zeta}$ ($= \prod_i \abs{\lambda_i}$) gives an additional factor 
for the probability distribution of the Hessian. 
Hence \eq{Hamiltonian} should be replaced by 
\beq
  H({\bm \lambda}) = \frac{1}{2} \lmk \sum_i \lambda_i^2 
  - \frac{a}{N} \lkk \sum_i \lambda_i \rkk^2  - \sum_{i \ne j} \ln \lmk \abs{\lambda_i - \lambda_j} \rmk 
  \rmk 
  - \sum_i \ln \abs{\lambda_i}. 
\label{Hamiltonian2}
\eeq

As we did in Sec.~\ref{sec:analytic1}, we rescale the eigenvalues as ${\bm \mu} = {\bm \lambda} / \sqrt{N}$ and consider a density function of ${\bm \mu}$. 
In terms of this density function, 
we can rewrite $H({\bm \lambda})$ as 
\beq
 H [ \rho] / N^2 &=& \frac{1}{2} \int \dd \mu \mu^2 \rho (\mu) 
 - \frac{1}{2} a \int \dd \mu \dd \mu' \rho(\mu) \rho(\mu') \mu \mu'
 \nonumber\\
 &&- \frac{1}{2} \int \dd \mu \dd \mu' \rho(\mu) \rho(\mu') \ln \lmk \abs{\mu - \mu'} \rmk 
 - \frac{1}{N} \int \dd \mu \rho (\mu) \ln \abs{\mu}. 
\eeq
The partition function $Z (\lambda_{\rm cr})$ can also be expressed in terms of $\rho (\mu)$, 
as in \eq{Z_cr}, where $\Sigma_0 [\rho]$ is given by \eq{Z_0}, while $\Sigma_1 [\rho]$ is now given by 
\beq
  \Sigma_1 [ \rho] &=& 
   \frac{1}{2} N(1-a) \int \dd \mu \dd \mu' \rho(\mu) \rho(\mu') \mu \mu' 
 + \int \dd \mu \rho (\mu) \ln \lkk \rho (\mu) \rkk 
 - \int \dd \mu \rho (\mu) \ln \abs{\mu}. 
 \nn
 \label{Sigma_1*}
\eeq

We can absorb the third term in (\ref{Sigma_1*}) into the first term in the following way. 
To the leading order in $N$, the distribution $\rho_c(\mu)$ at the saddle point is the shifted Wigner semi-circle
\beq
\rho_W(\mu,{\bar\mu})=\pi^{-1}\sqrt{2-\left(\mu-{\bar\mu}\right)^2}.
\label{Wigner semi-circle}
\eeq
Since $\Sigma_1 [\rho]$ is the next-leading order term, 
we can approximate it as $\Sigma_1 [\rho_{\rm W} (\mu ; \bar{\mu})]$. 
Then we can calculate the third term in (\ref{Sigma_1*}): 
\beq
 \int \dd \mu \rho_{\rm W} (\mu; \bar{\mu}) \ln \abs{\mu} 
 = \frac{\bar{\mu}^2}{2} - \frac{1}{2} \lmk 1 + \log 2 \rmk, 
\eeq
for $\abs{\bar{\mu} } \le \sqrt{2}$. 
Using the definition 
\beq
\bar{\mu} = \int \dd \mu \rho(\mu) \mu,
\eeq 
we can rewrite $\Sigma_1$ as 
\beq
  \Sigma_1 [ \rho] &\simeq& 
   \frac{1}{2} \lkk N(1-a) - 1 \rkk \int \dd \mu \dd \mu' \rho(\mu) \rho(\mu') \mu \mu' 
 + \int \dd \mu \rho (\mu) \ln \lkk \rho (\mu) \rkk 
 + ({\rm const.}), 
 \nn
 \label{sigma1:stationaryP}
\eeq
in the large $N$ limit. 
Therefore, we can use the result of the previous subsection, (\ref{x_0-2}) and (\ref{Sigma_c}),  with $a$ replaced by $a + 1/N$.

\subsection{Probability of minima and the smallest eigenvalue}
\label{sec:minima}

An important characteristic of a landscape is the density of potential minima in the field space.  If the correlation length of the landscape is $\Lambda$, the density of stationary points, where $\partial_i U=0$ is $\sim \Lambda^{-N}$ -- that is ${\cal O}(1)$ points per correlation volume.  The density of minima can be obtained by multiplying this by the probability for a stationary point to be a local minimum.  This is the same as the probability for all Hessian eigenvalues at that point to be positive,
\beq
P_{\rm min} = \exp(-N^2 \Delta \Sigma(0)) ,
\eeq
where $\Delta\Sigma (0)= \Sigma(\mu_{\rm cr}=0) -\Sigma(\mu_{\rm cr}=-\infty)$.  

\begin{figure}[t] %  figure placement: here, top, bottom, or page
   \centering
   \includegraphics[width=2.5in]{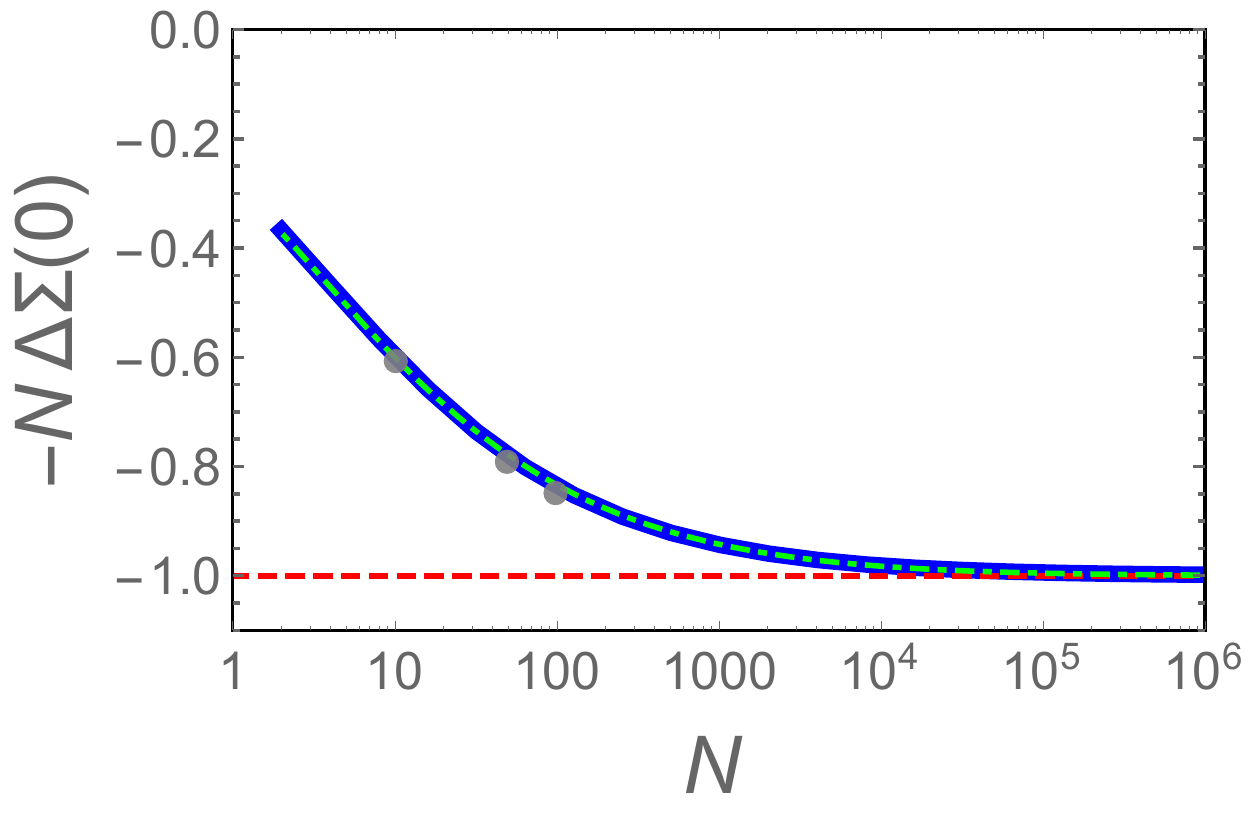} 
   \quad 
   \includegraphics[width=2.5in]{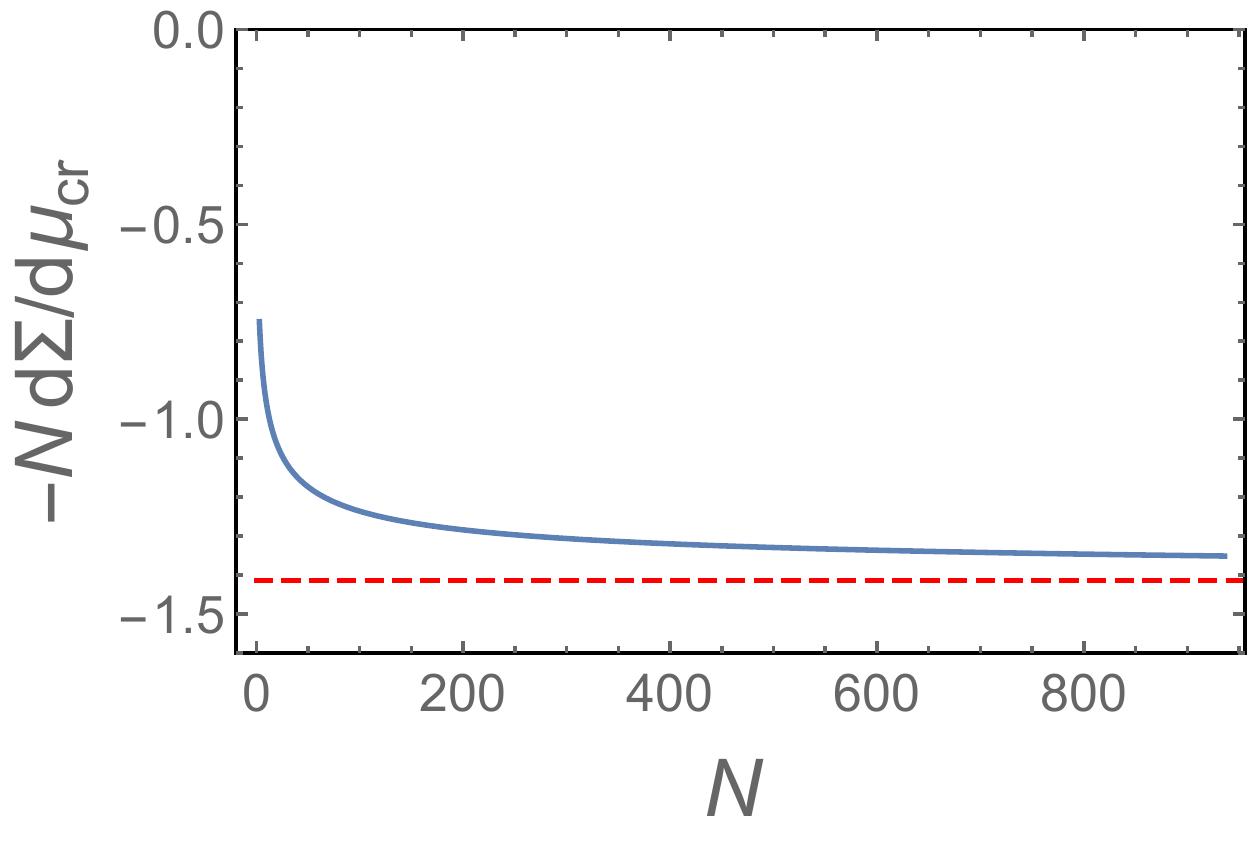} 
   \caption{
   {\bf Left:} 
$N \Delta \Sigma (0)$ as a function of $N$ for the Hessian distribution  
at stationary points of the potential (solid blue line).
The green dash-dotted line, which is completely overlapped with the result, 
is a fitting function given by~\eq{fit-sigma}. 
   The red dashed line marks the asymptotic value of $N \Delta \Sigma (0)$ in the limit of $N \to \infty$. 
   {\bf Right:} 
   $N d \Delta \Sigma (\mu_{\rm cr}) / d \mu_{\rm cr}$ at $\mu_{\rm cr} = 0$ 
   as a function of $N$. 
   We used $a = 1 - 1/N$.
   }
   \label{fig:sigma3}
\end{figure}

The left panel of Fig.~\ref{fig:sigma3} shows $N \Delta \Sigma (0)$ as a function of $N$ for the Hessian ensemble of \eq{Q final2}, where $a = N / (N+2)$ is replaced by $a + 1/N \simeq 1 - 1/N$. 
It gets close to the asymptotic value ($ - 1$) for $N \gtrsim 10^4$, but significantly deviates from that value at smaller values of $N$.  
The result can be well fitted by the following function, which is shown as the green dash-dotted line in the figure: 
\beq
 N \Delta \Sigma (0) \simeq 1 - 0.70 \, \exp \lkk -0.18 \, ( \ln N)^{1.36} \rkk. 
 \label{fit-sigma}
\eeq

The probability $P_{\rm min}$ has been studied earlier in the literature. 
Bray and Dean used the saddle point approximation in the large $N$ limit 
and found the asymptotic value $N \Delta \Sigma (0) = 1$~\cite{Bray:2007tf}. 
Easther et al~\cite{EastherGuthMasoumi} noted that $P_{\rm min}$ can significantly deviate from this value for moderately large values of $N$.  They calculated the probability using efficient numerical codes for several values of $N$ up to $100$.  Their results, shown by grey dots in the left panel of Fig.~\ref{fig:sigma3}, are in a very good agreement with ours.
However, their fitting formula is not consistent with ours at larger values of $N$.  
It is clear from the left panel of Fig.~\ref{fig:sigma3} that 
the asymptotic behavior cannot be correctly obtained by the extrapolation from $N \le 100$.

For some applications it is important to estimate the smallest eigenvalue of the Hessian at potential minima (see, e.g., Sec.~\ref{sec:second}).  The probability that this eigenvalue is greater than a given value $\mu_{\rm min}$ can be found from 
\beq
P_> (\mu_{\rm min}) = \exp(-N^2 \Delta \Sigma) ,
\eeq
where now $\Delta\Sigma = \Sigma(\mu_{\rm min}) -\Sigma(0)$.  
For small values of $\mu_{\rm min}$ we can approximate this as
\beq
P_>( \mu_{\rm min}) = 
 \exp \lmk \left. - N^2  \frac{d \Sigma (\mu_{\rm cr})}{d \mu_{\rm cr}} \right\vert_{\mu_{\rm cr} = 0} \mu_{\rm min} \rmk. 
\eeq
The probability distribution for the smallest eigenvalue can then be estimated as
\beq
{\cal P}( \mu_{\rm min}) = -\frac{d P_> (\mu_{\rm min})}{d \mu_{\rm min} }.
\eeq

We plot $N d \Delta \Sigma (\mu_{\rm cr}) / d \mu_{\rm cr}$ at $\mu_{\rm cr} = 0$ 
for the case of $a = 1 - 1/N$ in the right panel of Fig.~\ref{fig:sigma3}.  We see that it is $\sim 1$ for $N\sim 100$ and is asymptotic to $\sqrt{2}$ at $N \to \infty$, as shown by the red dashed line, 
in agreement with the analytic formula (\ref{sigma-approximation}). 
The typical magnitude of the smallest eigenvalue can now be estimated as 
\beq
 \mu_{\rm min} \sim \frac{1}{N^2 \left. \frac{d \Sigma (\mu_{\rm cr})}{d \mu_{\rm cr}} \right\vert_{\mu_{\rm cr} = 0}} 
 \sim \frac{1}{N}. 
 \label{mu_min:analytic}
\eeq

\subsection{Probability of $\mu > \mu_{\rm cr}$ for a fixed $U$}
\label{sec:fixedU}

We shall now use the distribution (\ref{Q final}) 
to calculate the probability for all Hessian eigenvalues to be greater than $\lambda_{\rm cr}$
at a given value of $U$.  Eq.~(\ref{Q final}) can be rewritten as
\beq
 Q_{U,\lambda} \simeq \frac{1}{4A} \lmk \sum_i (\lambda_{i} - \lambda_{*} )^2 -  \frac{1}{N} \lmk 1 -\frac{2 AE}{N (AE - B^2)} \rmk 
 \lkk \sum_i \lmk \lambda_{i} - \lambda_{*} \rmk \rkk^2 \rmk, 
 \nn
\eeq
where $\lambda_{*} = (B/E) (U - \bar{U})$ and we disregard terms that are independent of $\lambda_{i}$. 
We define shifted and rescaled eigenvalues ${\tilde\lambda}_i$ and the parameter $a$ as 
\beq
 &&{\tilde\lambda}_i = \frac{1}{\sqrt{2A}} \lmk \lambda_{i} - \lambda_{*} \rmk 
 \\
 &&a = 1 - \frac{2AE}{N (AE - B^2)}, 
 \label{shifted a}
\eeq
Then the resulting distribution for ${\tilde\lambda}_i$ has the same form as \eq{Q_zeta}. 

\begin{figure}[t] %  figure placement: here, top, bottom, or page
   \centering
   \includegraphics[width=4.5in]{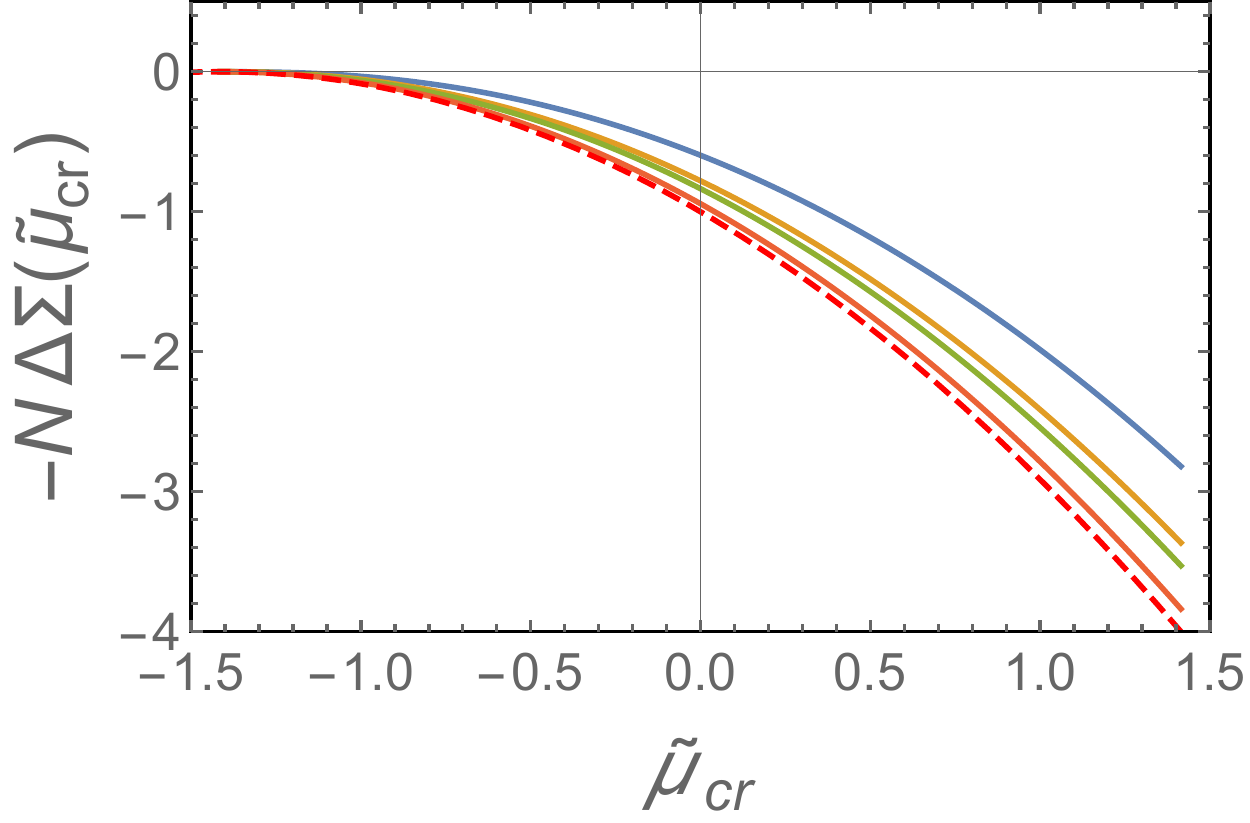} 
   \caption{
   $N \Delta \Sigma ({\tilde\mu}_{\rm cr})$ as a function of ${\tilde\mu}_{\rm cr}$. 
   We take $N = 10, 50, 100, 1000$ from top to bottom. 
   The dashed red line shows the asymptotic behavior at $N \to \infty$. 
   }
   \label{fig:sigma2}
\end{figure}

The condition $\lambda_i > \lambda_{\rm cr}$ is equivalent to ${\tilde\lambda}_i > {\tilde\lambda}_{\rm cr}$, where ${\tilde\lambda}_{\rm cr} = (1/\sqrt{2A}) (\lambda_{\rm cr} - \lambda_{*})$. 
Defining 
${\tilde\mu}_a = {\tilde\lambda}_a / \sqrt{N}$ ($a = i, {\rm cr}$), 
we obtain the relation between ${\tilde\mu}_{\rm cr}$ and 
the original threshold value $\lambda_{\rm cr}$: 
\beq
 {\tilde\mu}_{\rm cr} = \frac{1}{\sqrt{2AN}} \lmk \lambda_{\rm cr} - \lambda_{*} \rmk. 
 \label{shifted mu_cr}
\eeq
Thus we can use the same calculations and results as in Sec.~\ref{sec:analytic} 
for the conditional probability at generic points in the landscape, with the replacements (\ref{shifted a}) and (\ref{shifted mu_cr}). 
In particular, when we are interested in the case where all eigenvalues are positive, we should set 
$\lambda_{\rm cr} = 0$ or 
${\tilde\mu}_{\rm cr} = -\lambda_{*} (U) / \sqrt{2 A N}$. 
We plot $\Delta \Sigma ({\tilde\mu}_{\rm cr})$ as a function of ${\tilde\mu}_{\rm cr}$ for $N = 10, 50, 100, 1000$ 
and $(1-a) =1/N$ in Fig.~\ref{fig:sigma2}. 
The plots asymptote to the analytic formula (\ref{sigma-approximation}) 
or 
\beq
 N \Delta \Sigma ({\tilde\mu}_{\rm cr}) = 
  \frac{1}{2} \lmk {\tilde\mu}_{\rm cr} + \sqrt{2} \rmk^2 N (1-a), 
\label{deltaSigma:analytic}
\eeq
in the limit of $(1-a) \to 0$, which is plotted as a red dashed line.

To calculate the probability for a stationary point at a given value of $U$ to have all Hessian eigenvalues greater than $\lambda_{\rm cr}$, 
we need to add a term $- \sum_i \ln \abs{ \lambda_{i}}$ 
in \eq{Hamiltonian}. 
Neglecting a constant term, 
the additional term can be written as 
$-\sum_i \ln \abs{{\tilde\lambda}_i + \lambda_*}$. 
So we should replace $\lambda_i \to {\tilde\lambda}_i$ and
$-\sum_i \ln \abs{\lambda_i} \to -\sum_i \ln \abs{{\tilde\lambda}_i + \lambda_*}$ 
in \eq{Hamiltonian2}. 
By using the argument around \eq{Wigner semi-circle}, 
we can replace the extra term by  
\beq
  - \int \dd {\tilde\mu} \rho_{\rm W} ({\tilde\mu}; \bar{\mu}) \ln \abs{{\tilde\mu} + \mu_*} 
 &=& - \frac{(\bar{\mu} + \mu_*)^2}{2} + \frac{1}{2} \lmk 1 + \log 2 \rmk 
 \nn
 &=& 
 - \frac{1}{2} \int \dd {\tilde\mu} \dd {\tilde\mu}' {\tilde\rho} ({\tilde\mu}) {\tilde\rho} ({\tilde\mu}') {\tilde\mu} {\tilde\mu}' 
 - \mu_* \int \dd {\tilde\mu} {\tilde\rho} ({\tilde\mu}) {\tilde\mu} + ({\rm const.}), 
 \nn
\eeq
in the leading order in the large $N$ limit, 
where $\mu_* \equiv \lambda_{*} / \sqrt{2AN}$ 
and ${\tilde\rho} ({\tilde\mu}) = \rho (\mu)$. 
As a result, we can rewrite $\Sigma_1$ as 
\beq
  &&\Sigma_1 [ {\tilde\rho}] \simeq 
   \frac{1}{2} \lkk N(1-a) - 1 \rkk \int \dd {\tilde\mu} \dd {\tilde\mu}' {\tilde\rho}({\tilde\mu}) {\tilde\rho}({\tilde\mu}') 
   \lmk {\tilde\mu} - \frac{\mu_*}{N(1-a) - 1} \rmk \lmk {\tilde\mu}' - \frac{\mu_*}{N(1-a) - 1} \rmk 
   \nn
 &&~~~~~~~~~~~~~~~~+ \int \dd {\tilde\mu} {\tilde\rho} ({\tilde\mu}) \ln \lkk {\tilde\rho} ({\tilde\mu}) \rkk 
 + ({\rm const.}). 
 \nn
 \label{sigma1:stationaryP2}
\eeq
If we redefine ${\tilde\mu}$ by shifting ${\tilde\mu} \to {\tilde\mu} + \mu_* / [N(1-a) - 1]$, 
this is the same as \eq{sigma1:stationaryP} and we can use the same calculation and result. 
Therefore, the term $-\sum_i \ln \abs{\lambda_i + \lambda_*}$, which comes from the Jacobian 
for the stationary condition, gives two corrections: 
$a \to a + 1/N$ (as we discussed below \eq{sigma1:stationaryP})
and ${\tilde\mu} \to {\tilde\mu} +  \mu_* / [N(1-a) - 1]$.

In summary, 
the probability for a stationary point at a given value of $U$ to have all Hessian eigenvalues greater than $\lambda_{\rm cr}$ can be found from Eq.~(\ref{deltaSigma:analytic}) with the above replacements: 
\beq
N^2 \Delta\Sigma(\lambda_{\rm cr})=N\frac{AE +B^2}{2(AE-B^2)} 
 \lkk
 \frac{1}{\sqrt{2AN}} \lmk \lambda_{\rm cr} 
- \frac{2 B^2}{AE + B^2} \lambda_{*} \rmk + \sqrt{2}
\rkk^2 . 
\label{probmin}
\eeq

A similar result has been derived by Bray and Dean in their Eq.~(23) of Ref.~\cite{Bray:2007tf}, 
where their $f(0)$, $f'(0)$, $f''(0)$, $\epsilon$, are our $E/N$, $-B$, $AN$, $U / N$, respectively. 
There is, however, a significant difference.  Bray and Dean found the probability for a stationary point at a given value of $U$ to have a given index $\alpha$, where the index is defined as a fraction of eigenvalues which are negative.  The average eigenvalue ${\bar\lambda}$ in their Eq.~(23) has to be expressed in terms of $\alpha$ through the relation
\beq
\int_{-\infty}^0 d\lambda \rho(\lambda,{\bar\lambda}) = \alpha.
\eeq
Their $\bar{\lambda}$ can be identified with our $(\lambda_{\rm cr} + 2\sqrt{AN})$ in the leading order approximation.  We note, however, that Bray and Dean calculated $\rho(\lambda,{\bar\lambda})$ only in the leading order in $1/N$, which becomes rather inaccurate near the left edge of the distribution.  Hence their result is not accurate for small values of $\alpha$. 
We do not have this problem in our calculation, so our result can be used for arbitrary values of $\lambda_{\rm cr}$.

We note also that even though the approximations we used here are sufficient for calculating
$\Delta\Sigma(\lambda_{\rm cr})$, they are not accurate enough to find the distribution $\rho_c (\mu)$ at small values of $\mu$, because the distribution strongly deviates from the Wigner semi-circle near $\mu=0$.  In Sec.~\ref{sec:dynamical} we shall develop a new method which is sufficiently accurate in that regime.

\section{Dynamical method} 
\label{sec:dynamical}

As we already noted, the approximations we used in Sections \ref{sec:analytic2} and \ref{sec:minima} are not sufficiently accurate for finding the eigenvalue distribution at small values of $\mu$.  In this section we develop a numerical method, using a version of the Dyson Brownian motion \cite{Dyson:1962}, 
to dynamically derive the distribution of Hessian eigenvalues.  This method accounts for all terms in $\Sigma_0$ and $\Sigma_1$ without any approximations, apart from the limitations of numerical resolution and computer runtime.  
We shall first apply the Dyson Brownian motion method to the random matrix theory with GOE.

\subsection{Random matrix theory}

\subsubsection{Dyson Brownian motion and Fokker-Planck equation} 

The Dyson Brownian motion model was introduced in Ref.~\cite{Dyson:1962} to describe stochastic evolution of random matrices.  The eigenvalues $\lambda_i$ of a random matrix are assumed to undergo a stochastic process described by the Langevin equation
\beq
 \frac{\dd \lambda_i (t)}{\dd t} = - \frac{\del W}{\del \lambda_i} 
 + \xi_i (t), 
 \label{EoM of lambda}
\eeq
where $\xi_i (t)$ is a stochastic variable, 
\beq
 \la \xi_i (t) \xi_j (t') \ra = 2 \delta_{ij} \delta ( t - t'). 
\eeq
The potential $W$ is given by 
\beq
 W = \frac{1}{2} \sum_i \lambda_i^2 - \frac{1}{2} \sum_{i \ne j} \ln \abs{\lambda_i - \lambda_j}. 
\label{W}
\eeq
The eigenvalues are subject to a potential force $\del W / \del \lambda_i$ and a stochastic force $\xi_i$.  Note that the potential $W$ is equal to the `Hamiltonian'  (\ref{Hamiltonian}). 

The probability density $P({\bm \lambda}, t)$ satisfies the Fokker-Planck equation, which can be obtained by taking the ensemble average over $\xi_i$ \cite{Dyson:1962}: 
\beq
 &&\frac{\del}{\del t} P({\bm \lambda}, t) 
 = - \sum_i \frac{\del }{\del \lambda_i}  j_i ({\bm \lambda}, t)
 \label{FPeq1}
 \\
 && j_i ({\bm \lambda}, t) = 
 -T' \frac{\del P }{\del \lambda_i} + 
 E_i P, 
\eeq
where $T' = 1$ and the potential force $E_i$ is given by 
\beq
 E_i \equiv - \frac{\del W }{\del \lambda_i}. 
\eeq

The equilibrium solution of \eq{FPeq1} is given by the Boltzmann distribution, 
\beq
P \propto \exp[ - W / T']. 
\label{Boltzmann}
\eeq
Thus we can interpret $W$ and $T'$ as the potential and the temperature, respectively.  We note that the distribution (\ref{Boltzmann}) is the same as the eigenvalue distribution (\ref{Boltz}), (\ref{Hamiltonian}) with $a=0$ for the GOE ensemble.  

Since we are interested not in the individual variables $\lambda_i$, but in the distribution of 
eigenvalues, we define a time-dependent probability density $\rho( \lambda, t)$ as 
\beq
 \rho (\lambda, t) = 
 \int \prod_i \dd \lambda_i 
 \lmk \frac{1}{N} \sum_i \delta \lmk \lambda - \lambda_i \rmk \rmk
 P(\lambda_1, \lambda_2, \dots, \lambda_N, t). 
 \label{Ptorho}
\eeq
We can easily check that $\int \dd \lambda \rho = 1$. 
We also rescale the variable as $\mu = \lambda /\sqrt{N}$ 
to compare the results with those in Sec.~\ref{sec:analytic}. 
We set the normalization condition $\int \dd \mu \rho (\mu, t) = 1$ 
so we rescale the density $\rho (\lambda , t ) \to  \rho (\mu, t)/ \sqrt{N}$. 
Then it obeys the following equation: 
\beq
 &&\frac{\del \rho (\mu, t)}{\del t} = - \frac{\del j (\mu, t)}{\del \mu} 
 \label{FPeq2}
 \\
 &&j(\mu, t) = 
 -T \frac{\del \rho}{\del \mu} + 
 E \rho, 
\label{j2}
\eeq
where $T = 1/N$ is the temperature.  The potential force $E$ is given by 
\beq
 E(\mu,t) 
= - \mu + \int \dd \mu' \frac{\rho(\mu',t)}{\mu - \mu'}. 
\eeq
where we have replaced the summation $\sum_j$ by the integral $\int \dd \mu' \rho (\mu')$ 
in the second term. 

In what follows we shall use the Fokker-Planck equation for $\rho(\mu,t)$, without referring to the Langevin equation.

\subsubsection{Dynamical evolution}
\label{sec:dynamicalGOE}

We are interested in the equilibrium distribution under the condition that all eigenvalues are positive. This can be realized by evolving $\rho (\mu, t)$ by \eq{FPeq2} for a sufficiently long time\footnote{We note that the equilibrium eigenvalue distribution in a different class of models has been studied in Ref.~\cite{Bachlechner:2012at}, where they calculated the distribution by sampling the canonical ensemble with the Metropolis algorithm.} with a reflecting boundary condition at $\mu=0$, 
\beq
j (\mu = 0 , t) = 0. 
\label{bc}
\eeq

The equilibrium solution is stationary, $\del \rho / \del t = 0$, and it follows from Eq.~(\ref{FPeq2}) that $j(\mu)={\rm const}$.  Then the boundary condition (\ref{bc}) requires that 
\beq
j(\mu)=-\frac{1}{N}\frac{\partial\rho}{\partial\mu}+E\rho =0
\label{j=0}
\eeq
This condition is similar to the one that we used to determine the saddle point solution 
(see \eq{saddle point2}): 
\beq
 - \frac{1}{N \rho(\mu)} \frac{\dd \rho_c (\mu)}{\dd \mu} 
 - \mu 
+\mathcal{P} \int_{\mu_{\rm cr}}^\infty \dd \mu' \frac{\rho_{\rm c} (\mu')}{\mu - \mu'} 
= 0, 
 \label{saddle point3}
\eeq
where we set $a = 0$.  The first term in Eq.~(\ref{saddle point3}) comes from the term $\int \dd \mu \rho \ln [ \rho]$ in $\Sigma_1[\rho]$, which we neglected in Sec.\ref{sec:analytic}.
Thus \eq{FPeq2} provides a useful check for the results of saddle point approximation.

\begin{figure}[t] %  figure placement: here, top, bottom, or page
   \centering
   \includegraphics[width=2.5in]{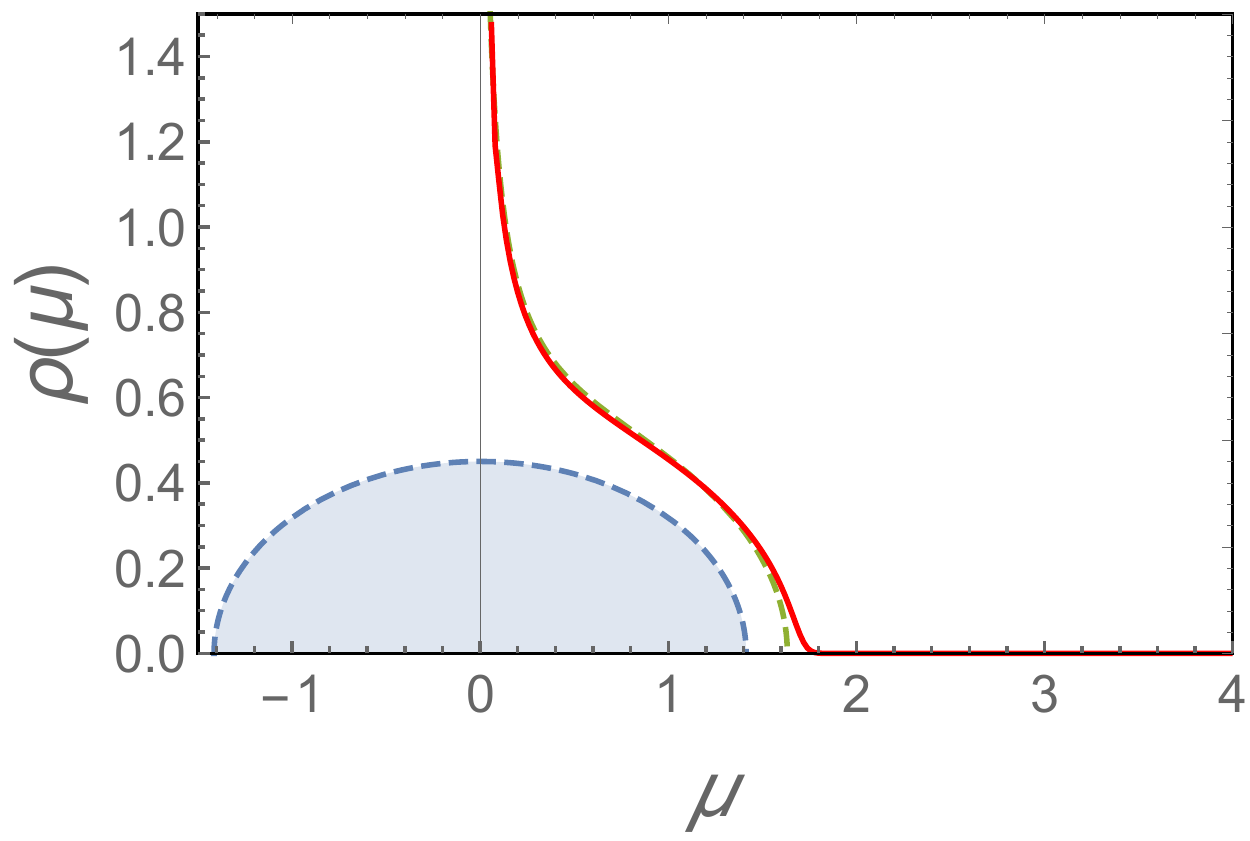} 
   \quad 
   \includegraphics[width=2.5in]{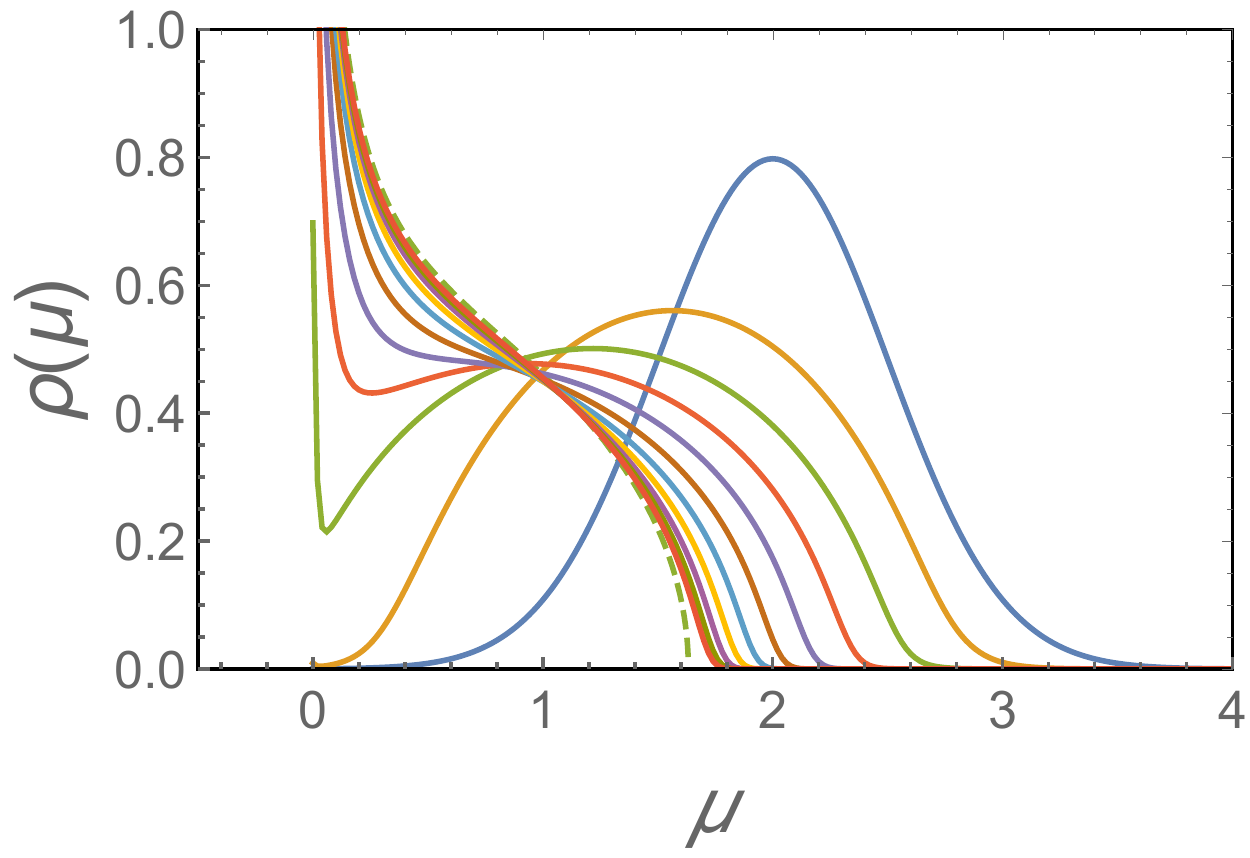} 
   \caption{
{\bf Left:} Equilibrium solution of Fokker-Planck equation for the GOE model (red curve). 
We also plot the Wigner semi-circle (dashed shaded blue line) and the analytic solution for the GOE model (dashed green line). 
{\bf Right:} Time-evolution of the distribution. We plot distributions at $t = 0.25 n$ with $n = 0,1,2, \dots, 10$.  The initial distribution is shown by a bell-shaped blue line.
   }
   \label{fig:FP-GOE}
\end{figure}

We solve the Fokker-Planck equation numerically by discretizing the differential equation. 
Numerical methods for solving the Fokker-Planck equation with the boundary condition (\ref{bc}) have been extensively studied~\cite{Chang,Mohammadi, Pareschi}. 
The grid size $\Delta \mu$, the volume of $\mu$-space $L_{\rm \mu}$, 
and the step size $\Delta t$ are taken to be $0.02$, $5$, and $0.005$, respectively. 
We checked that our results are not affected by these parameters by varying their values. 
The results are presented in Fig.~\ref{fig:FP-GOE}, where we take $N = 1/T = 100$. 
The initial condition is taken to be a Gaussian function with a peak at $\mu = 2$ and a width of $0.5$, as indicated by a blue line.  We see that the evolution converges to a stationary
distribution, which agrees very well with the semi-analytic solution of Sec.\ref{sec:analytic}, with only a slight deviation at the right edge.

Here we comment on this slight deviation.  It comes from the fact that we neglected the second term of $\Sigma_1[\rho]$ in \eq{Sigma1} to calculate the semi-analytic solution 
while we do not use any approximation to calculate $\rho$ in the dynamical method. 
To check that this deviation is physical and is consistent with the results in the literature, 
we calculated the distribution for the case of $\mu_{\rm cr} \ll - \sqrt{2}$ (i.e., for the case without the boundary) using the dynamical method.  We found that the tails of the distribution at the right and left edges agree very well with the well-known Tracy-Widom distribution \cite{Tracy:1992rf, Tracy:1995xi}. Therefore, the smooth tail of the distribution at the right edge in Fig.~\ref{fig:FP-GOE} can be attributed to the spread of the largest eigenvalues {\it a la} Tracy-Widom beyond the edge of the semi-analytic distribution.

\subsection{Hessian eigenvalue distribution in RGF model}
\label{sec:dynamical2}

The same method can be applied to find the Hessian eigenvalue distribution in a random Gaussian field, except in this case we should use $a=N/(N+2)$ in Eq.~(\ref{Hamiltonian}).
Then the potential (\ref{W}) is replaced by 
\beq
 W = \frac{1}{2} \sum_i \lambda_i^2 
 - \frac{a}{2N} \lmk \sum_i \lambda_i \rmk^2 
 - \frac{1}{2} \sum_{i \ne j} \ln \abs{\lambda_i - \lambda_j}, 
 \label{W2}
\eeq
and the potential force in the Fokker-Planck equation becomes
\beq
 E(\mu,t) =  - \mu  + a \int \dd \mu' \rho (\mu',t) \mu' 
 + \int \dd \mu' \frac{\rho(\mu',t)}{\mu - \mu'}
\label{E}
\eeq
The equilibrium distribution $\rho_{\rm c} (\mu)$ 
is again equivalent to the saddle point solution of (\ref{saddle point2}) 
with the term coming from $\int \dd \mu \rho \ln [\rho]$ included. 
We find this distribution by evolving $\rho (\mu,t)$ via the Fokker-Planck equation.

\begin{figure}[t] %  figure placement: here, top, bottom, or page
   \centering
   \includegraphics[width=2.5in]{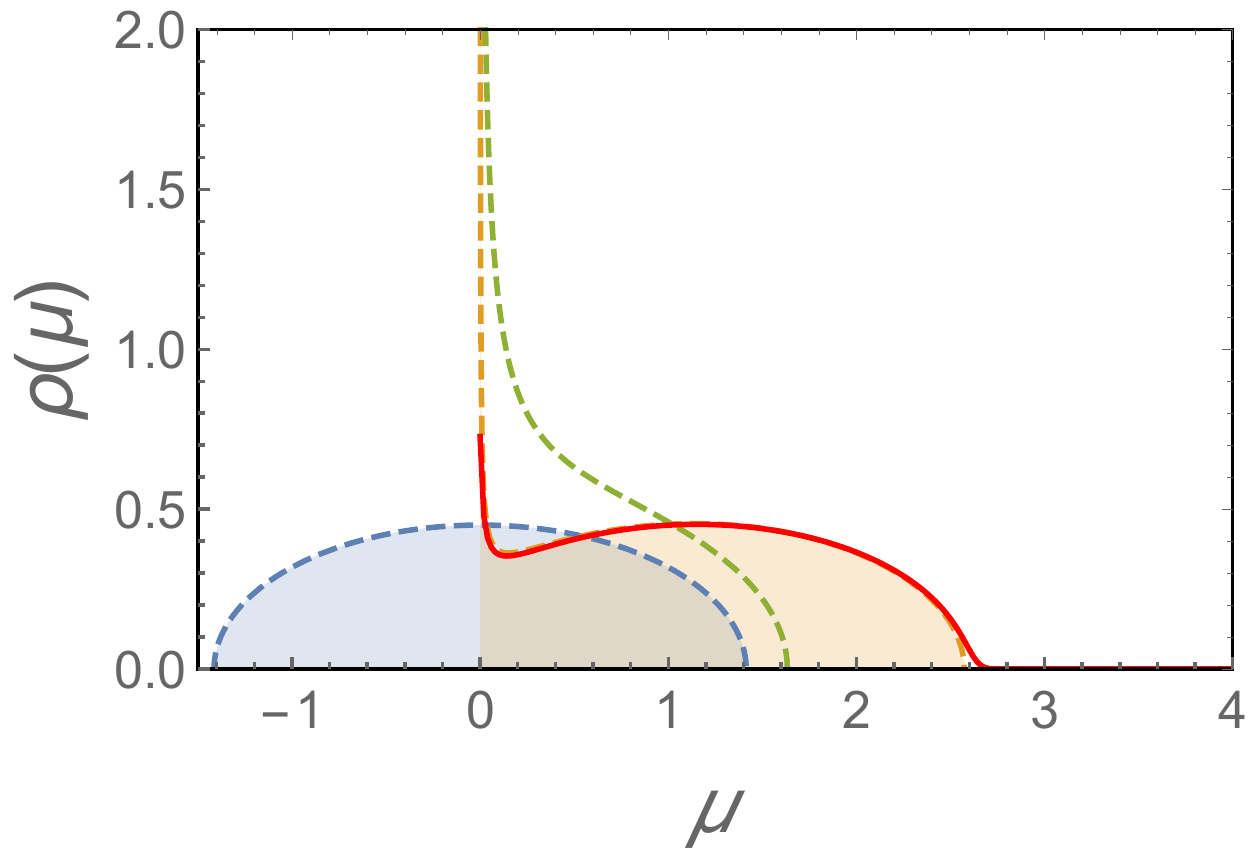} 
   \quad 
   \includegraphics[width=2.5in]{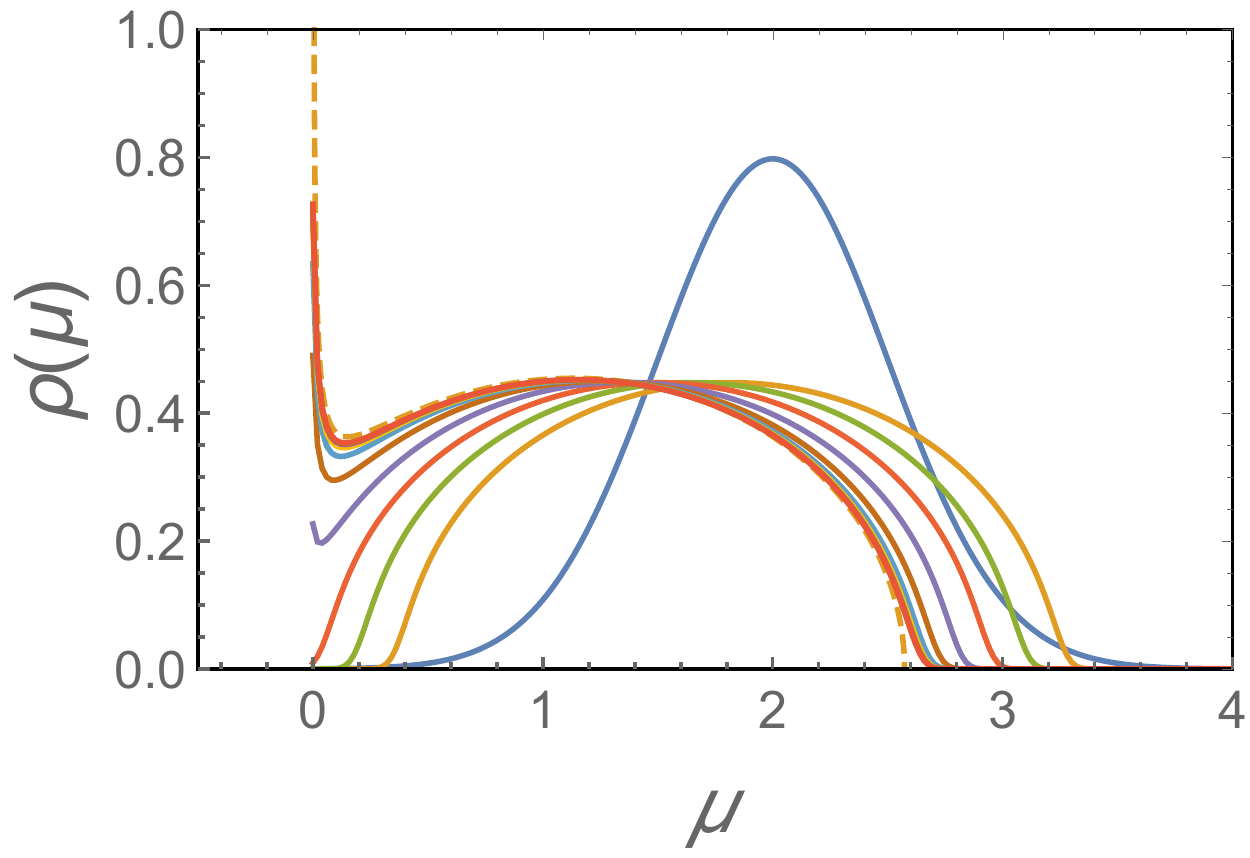} 
   \caption{
{\bf Left:} Equilibrium solution of Fokker-Planck equation for Hessian eigenvalue distribution at a generic point in RGF (shaded solid red line). 
We plot also the Wigner semi-circle and the analytic solutions for the GOE and RGF models (dashed green and orange lines, respectively). 
{\bf Right:} Time-evolution of the distribution. We plot distributions at $t = 5 n$ with $n = 0,1,2, \dots, 10$. 
   }
   \label{fig:FP-RGF}
\end{figure}

We solve the Fokker-Planck equation numerically and show the result in Fig.~\ref{fig:FP-RGF}. 
We take $N = 1/T = 100$ and $a = N/(N+2)$. 
The initial condition and other parameters are the same as we used in Sec.\ref{sec:dynamicalGOE} for the case of GOE.  Once again, we see that the endpoint of the evolution is very close to the analytic solution. This justifies the approximation of neglecting the 
term $\int \dd \mu \rho \ln [ \rho]$ in $\Sigma_1[\rho]$ that we made in Sec.~\ref{sec:analytic2}.

\subsection{Hessian eigenvalue distribution at stationary points of the potential}
\label{sec:dynamical2-2}

We finally consider the Hessian eigenvalue distribution at stationary points, where $\del_i U = 0$, under the condition that all eigenvalues are positive.  We found in Sec.\ref{sec:analytic2} that in this case $\Sigma_1 [\rho]$ has an additional term, $-\int \rho(\mu) \ln |\mu|$.  This adds an extra term $1/N \mu$ to the potential force (\ref{E}) in the Fokker-Planck equation, 
\beq
 E(\mu,t) =  - \mu  + a \int \dd \mu' \rho (\mu',t) \mu' 
 + \int \dd \mu' \frac{\rho(\mu',t)}{\mu - \mu'} + \frac{1}{N \mu} .
 \label{E2}
\eeq

\begin{figure}[t] %  figure placement: here, top, bottom, or page
   \centering
   \includegraphics[width=2.5in]{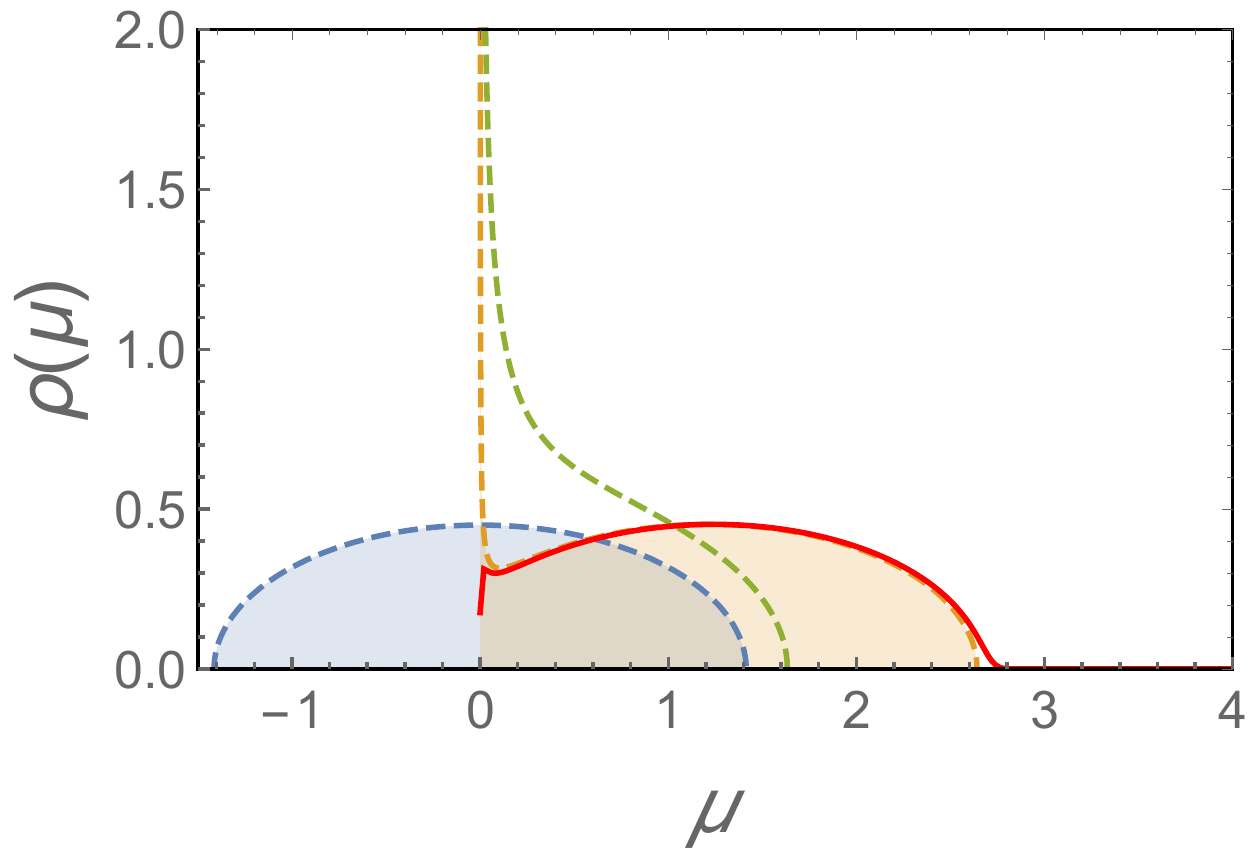} 
   \quad 
   \includegraphics[width=2.5in]{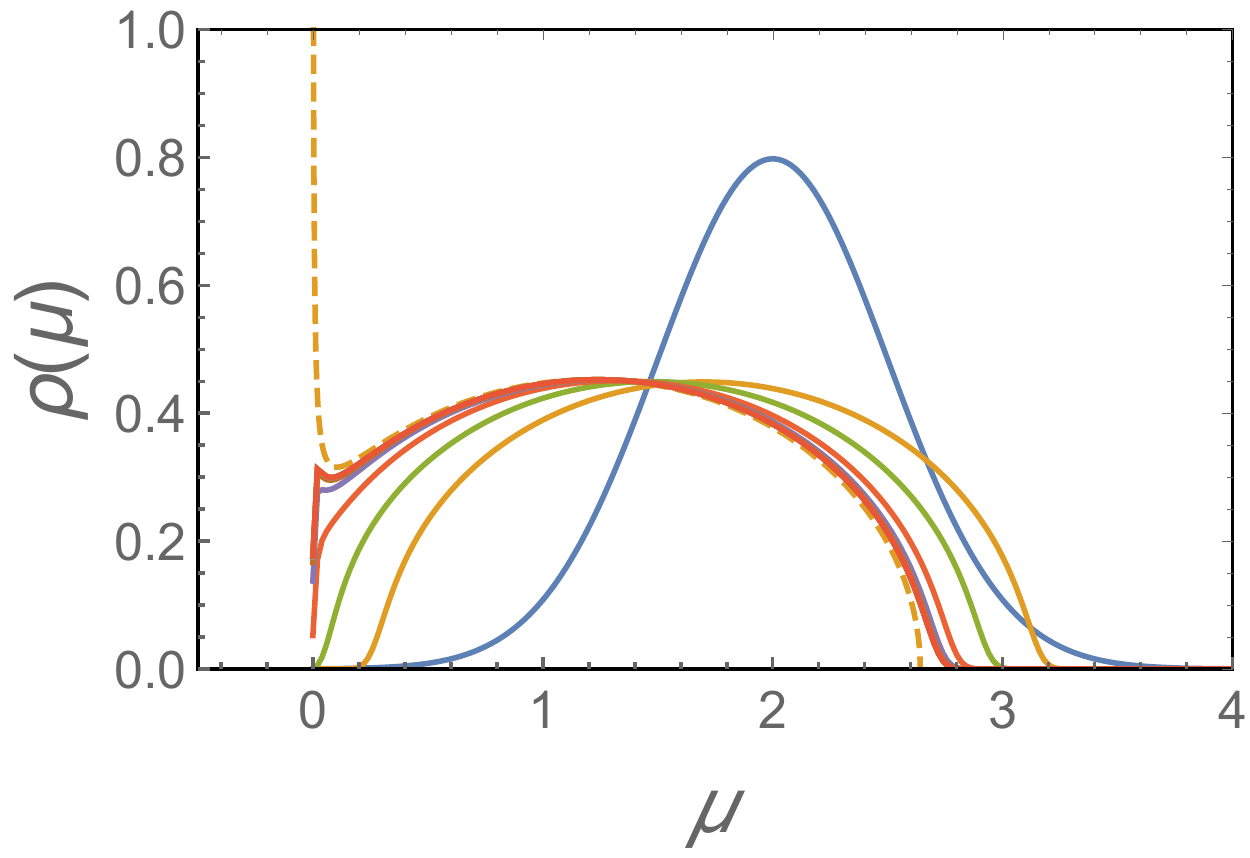} 
   \caption{
   {\bf Left:} Equilibrium solution of Fokker-Planck equation for Hessian eigenvalue distribution in RGF at local minima of the potential (solid red shaded line).  We plot also the Wigner semi-circle 
and analytic solutions for the GOE and RGF model (dashed green and orange lines, respectively). 
{\bf Right:} Time-evolution of the distribution. We plot distributions at $t = 10 n$ with $n = 0,1,2, \dots, 10$. 
   }
   \label{fig:FP-RGF2}
\end{figure}

We solve the equation numerically using the same parameter values and initial condition as before. The results are presented in Fig.~\ref{fig:FP-RGF2}. The equilibrium distribution is shown by the solid shaded red line.  For comparison we also show, by a dashed orange line, the semi-analytic distribution calculated in Sec.~\ref{sec:analytic2} for Hessian eigenvalues at generic points (not necessarily potential minima). We see that the two distributions are very close to one another, except near $\mu=0$.  The semi-analytic solution diverges as $\mu^{-1/2}$, while our equilibrium distribution drops sharply to zero.  This is the effect of the strong repulsive force due to the last term in Eq.~(\ref{E2}). 

\begin{figure}[t] %  figure placement: here, top, bottom, or page
   \centering
   \includegraphics[width=4.5in]{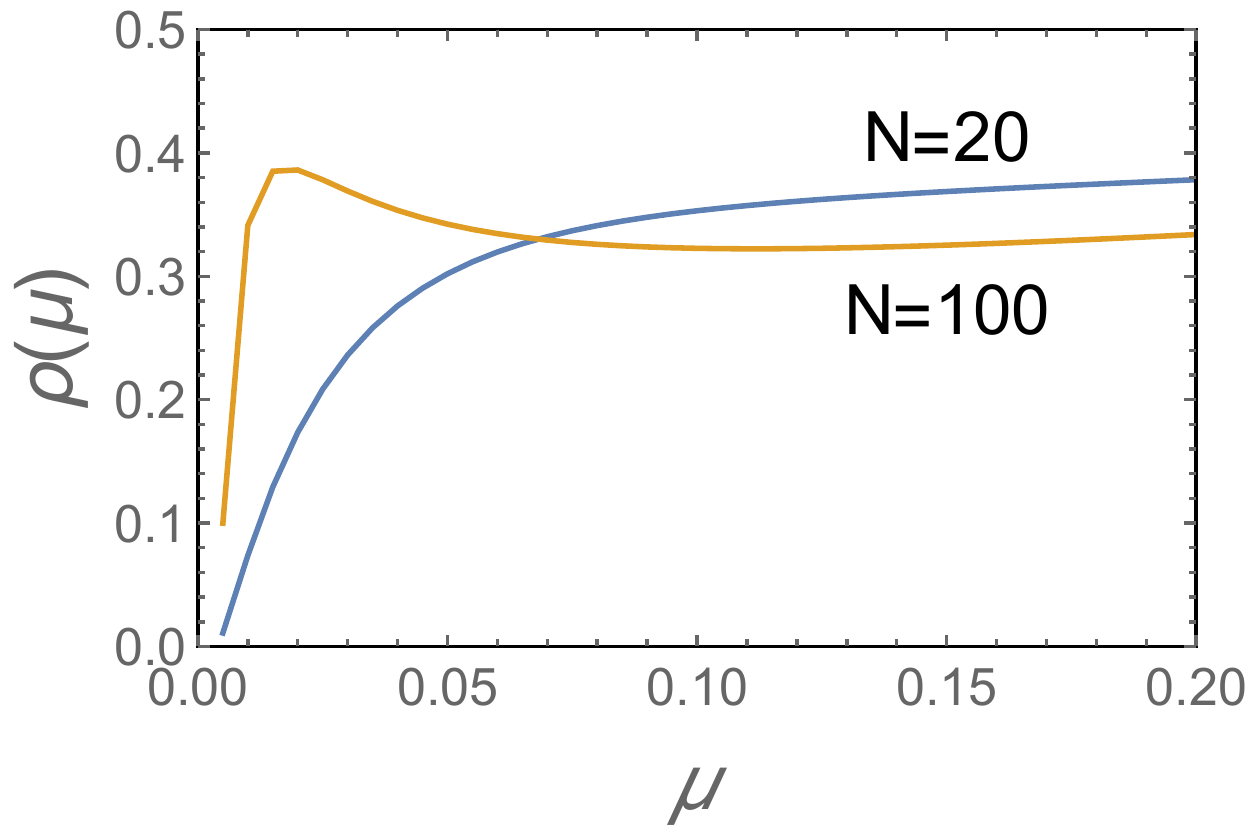} 
   \caption{
   Hessian eigenvalue distribution for small values of $\mu$ in RGF at local minima of the potential. 
   We plot the cases of $N = 20$ (blue line) and $100$ (yellow line). 
}
   \label{fig:FP-RGF-smallmu}
\end{figure}

To illustrate the behavior of the distribution at small values of $\mu$,  we plot $\rho_c(\mu)$ near $\mu = 0$ for the cases of $N = 20$ (blue line) and $100$ (yellow line) in Fig.~\ref{fig:FP-RGF-smallmu}.  For $\mu\ll 1/N$, we can approximate $E(\mu)\approx 1/N\mu$, and Eq.(\ref{j=0}) gives
\beq
\rho_c(\mu)\approx C\mu ~~~~ (\mu\ll 1/N)
\label{Cmu}
\eeq
with $C={\rm const}$.  This is in agreement with the plots in Fig.~\ref{fig:FP-RGF-smallmu}.
We find that $C$ is about $0.5 N$ from our numerical results.  

It should be noted, however, that our method may not be accurate in the range $0<\mu \lesssim 1/N$.  The average number of eigenvalues in this range is $N\int_0^{1/N} \rho_c(\mu)d\mu \sim 1$, and thus replacing discrete eigenvalues by a continuous distribution is not justified.\footnote{We believe, however, that the sharp drop of the distribution to zero at $\mu=0$ is a real feature.  A similar feature was found in Refs.~\cite{MarcenkoPastur} and \cite{Bachlechner:2012at}, where the eigenvalue distribution was calculated for different models without using the continuous approximation.}
One can expect nevertheless that this approximation gives correct order-of-magnitude results near the limit of its applicability, $\mu\sim 1/N$.  We can then use it to estimate the typical magnitude of the smallest eigenvalue of the Hessian, $\mu_{\rm min}$: 
\beq
\int_0^{\mu_{\rm min}} \rho_c(\mu)d\mu \sim \frac{1}{N}.
\label{mumineq}
\eeq
The plots in Fig.~\ref{fig:FP-RGF-smallmu} suggests that $\rho_c (\mu)\simeq 0.3$ for $\mu\gtrsim 1/N$.  Hence we find
\beq
 \mu_{\rm min} \sim \frac{1}{N}. 
 \label{mu_min}
\eeq
The same estimate is obtained by numerically integrating the distribution in Eq.~(\ref{mumineq}).
It is in agreement with a more accurate estimate (\ref{mu_min:analytic}) in Sec.~\ref{sec:minima}.

\section{Some applications in cosmology}
\label{sec:implication}

In this section we consider some applications of our result to the landscape models.

\subsection{Vacuum stability}

Vacuum stability in landscape models has been studied numerically in Refs.~\cite{Greene:2013ida,MV}. A simple analytic treatment was given by Dine and Paban in Ref.~\cite{Dine:2015ioa}.  They assume {\it (i)} that the most probable decay channels are typically in the directions of the smallest Hessian eigenvalues and {\it (ii)} that the vacuum decay rate is controlled mainly by the quadratic and cubic terms in the expansion of $U({\bf \phi})$ about the potential minimum.  Then the tunneling (bounce) action in the direction of the Hessian eigenvalue $\lambda_i$ can be estimated as
\beq
B_i \sim K\frac{\lambda_i}{\gamma^2},
\eeq
where $\gamma\sim U_0/\Lambda^3$ is the typical coefficient of a cubic expansion term and $K\sim 50$ is a numerical coefficient.  The highest rate corresponds to the smallest Hessian eigenvalue $\lambda_{\rm min}$.  Dine and Paban assume that $\lambda_{\rm min} \sim (1/N) (U_0/\Lambda^2)$ and find $B\sim (K/N) (\Lambda^4/U_0)$.  With $U_0/\Lambda^4 \sim 0.1-1$ and $N\sim 100$, this can be rather small, $B\sim 1$, suggesting that most of the vacua in the landscape are very unstable.

A more accurate estimate of $\lambda_{\rm min}$ can be obtained from \eq{mu_min:analytic} or Eq.~(\ref{mu_min}), 
\beq
 \mu_{\rm min} \sim \frac{1}{N}. 
\eeq
This corresponds to
\beq
\lambda_{\rm min} \sim \frac{U_0\sqrt{N}}{\Lambda^2} \mu_{\rm min} \sim \frac{U_0}{\Lambda^2 \sqrt{N}},
\eeq
and thus the tunneling action is 
\beq
B\sim \frac{K}{\sqrt{N}} \frac{\Lambda^4}{U_0}.
\eeq
This is $\sqrt{N}$ times larger than the estimate of Ref.~\cite{Dine:2015ioa}, so the vacuum stability is significantly enhanced.  (We note that if we used Eq.~(\ref{mumineq}) with the GOE eigenvalue distribution (\eq{rho_c} with $x_0 = 0$), we would have $\mu_{\rm min}\sim 1/N^2$, which would suggest a much lower stability.)

\subsection{Multi-field inflation}
\label{sec:second}

In this Section we assume that the landscape is small-field, which means that the correlation length is $\Lambda\ll 1$ in Planck units.  Slow-roll inflation in such a landscape occurs in rare flat regions, where the first and second derivatives of the potential in some direction are much smaller than their typical values.  It was argued in Refs.~\cite{MVY1, MVY2, MVY3, Blanco-Pillado:2017nin} that inflation in such regions tends to be single-field, with the inflaton field rolling in a nearly straight line along the flat direction.  Other fields (corresponding to orthogonal directions) can be excited and significant deviations from a straight trajectory can occur only if some of the fields have masses smaller than the Hubble parameter during inflation, $m\lesssim \sqrt{U_0}$ in Planck units.  This is much smaller than the typical mass $m_0 \sim\sqrt{U_0}/\Lambda$.  However, with a large number of fields $N$ some of the masses may be $\ll m_0$ and may get as small as $\sqrt{U_0}$.  We shall now investigate this possibility.

Flat inflationary tracks are likely to be found in the vicinity of inflection points, where one of the Hessian eigenvalues vanishes (this corresponds to the flat direction), the rest of the eigenvalues are positive, and the potential gradient vanishes in the directions orthogonal to the flat direction.  Let us choose the $\phi_1$ axis in the flat direction.  Then we have $\lambda_1=0$ and $\lambda_i >0$, $\del U / \del \phi_i =0$ for $i=2, ..., N$.  The mass spectrum in the directions orthogonal to the flat direction is determined by the Hessian eigenvalues, $m_i^2 = \lambda_i ~ (i>1)$.  We now want to estimate the smallest of these eigenvalues. 

The probability distribution for Hessian eigenvalues ${\bm \lambda} = ( \lambda_2, \lambda_3, \dots, \lambda_N)$ at inflection points can be derived along the same lines as we derived \eq{Hamiltonian2}.  It is given by 
\beq
 &&P =A \exp ( - H ({\bm \lambda}) 
 \\
 &&H ({\bm \lambda}) 
 = \frac{1}{2} \lmk \sum_{i \ge 2} \lambda_i^2 
  - \frac{a}{N} \lkk \sum_{i \ge 2} \lambda_i \rkk^2  - 2 \sum_{i > j \ge 2} \ln \lmk \abs{\lambda_i - \lambda_j} \rmk 
  \rmk 
  - 2 \sum_{i \ge 2} \ln \abs{\lambda_i}. 
  \label{Hamiltonian4}
\eeq
This is similar to \eq{Hamiltonian2}, but with a few differences. 
First, the coefficient of the last term is not unity but is $2$.  An additional $-\sum \ln |\lambda_i|$ term comes from the last term in parentheses of Eq.~(\ref{Hamiltonian2}) with $i=1$ or $j=1$.
Second, the number of eigenvalues is $N-1$. 

We can now use the method of Sec.~\ref{sec:minima} to find the probability distribution for the second smallest Hessian eigenvalue $\mu_2$ at inflection points (the first smallest being $\mu_1=0$). 
We note that the number of eigenvalues is now $N-1$ and rewrite the coefficient of the second term in the parenthesis of (\ref{Hamiltonian4}) as $a/N=a'/(N-1)$, where $a' = a (N-1)/ N$, so this term becomes 
\beq
- \frac{a'}{N-1} \lkk \sum_{i \ge 2} \lambda_i \rkk^2. 
\eeq
As we explained in Sec.~\ref{sec:analytic2}, 
the last term of \eq{Hamiltonian4} 
can be absorbed into $a'$ 
by the replacement of $a' \to a' + 2/N$, 
where the factor of $2$ comes from the coefficient of the last term. 
As a result, we should replace $a$ with 
$a (N-1)/ N + 2 /N$ in the calculation of Sec.~\ref{sec:analytic1}. 
Since 
$a (N-1)/ N + 2 /N \simeq a + 1/N$, 
the result should be the same with the one obtained in Sec.~\ref{sec:analytic2} 
in the large $N$ limit. 
Therefore the distribution of eigenvalues at an inflection point 
is given by $\rho_c(\mu)$ 
with $1 - a \simeq 1/N$ and $\mu_{\rm cr} = 0$. 
The probability for all eigenvalues to be positive is given by $\exp [ - N^2 \Delta \Sigma]$, and the typical value of $\mu_2$ can be estimated as in \eq{mu_min:analytic}, 
\beq
 \mu_2 \sim \frac{1}{N^2 \left. \frac{d \Sigma (\mu_{\rm cr})}{d \mu_{\rm cr}} \right\vert_{\mu_{\rm cr} = 0}} 
 \sim \frac{1}{N}. 
 \label{mu_2:analytic}
\eeq
The asymptotic value of $N d \Sigma (\mu_{\rm cr}) / d \mu_{\rm cr} \vert_{\mu_{\rm cr} = 0}$ 
is $\sqrt{2}$ in the limit $N \to \infty$.

The distribution of eigenvalues at inflection points can be found using the dynamical method of Sec.~\ref{sec:dynamical}. 
The Fokker-Planck equation has the same form as before, but with 
slightly different parameters and coefficients. 
The temperature $T$ in \eq{j2} is given by $1/(N-1)$ 
and 
the potential force $E(\mu,t)$ is given by 
\beq
 E(\mu,t) =  - \mu  + \frac{a (N-1)}{N} \int \dd \mu' \rho (\mu',t) \mu' 
 + \int \dd \mu' \frac{\rho(\mu',t)}{\mu - \mu'} + \frac{2}{(N-1) \mu}, 
\eeq
where $\mu \equiv \lambda / \sqrt{N-1}$. 
The change in the last term of $E(\mu,t)$ 
modifies the form of the distribution at $\mu \to 0$.  In this limit, the Fokker-Planck equation reduces to $d\rho/d\mu =2\rho/\mu$, with the solution 
\beq
\rho_c(\mu)\approx C\mu^2 ~~~~ (\mu\ll 1/N)
\label{Cmu^2}
\eeq
where $C={\rm const}$.  

\begin{figure}[t] %  figure placement: here, top, bottom, or page
   \centering
   \includegraphics[width=4.5in]{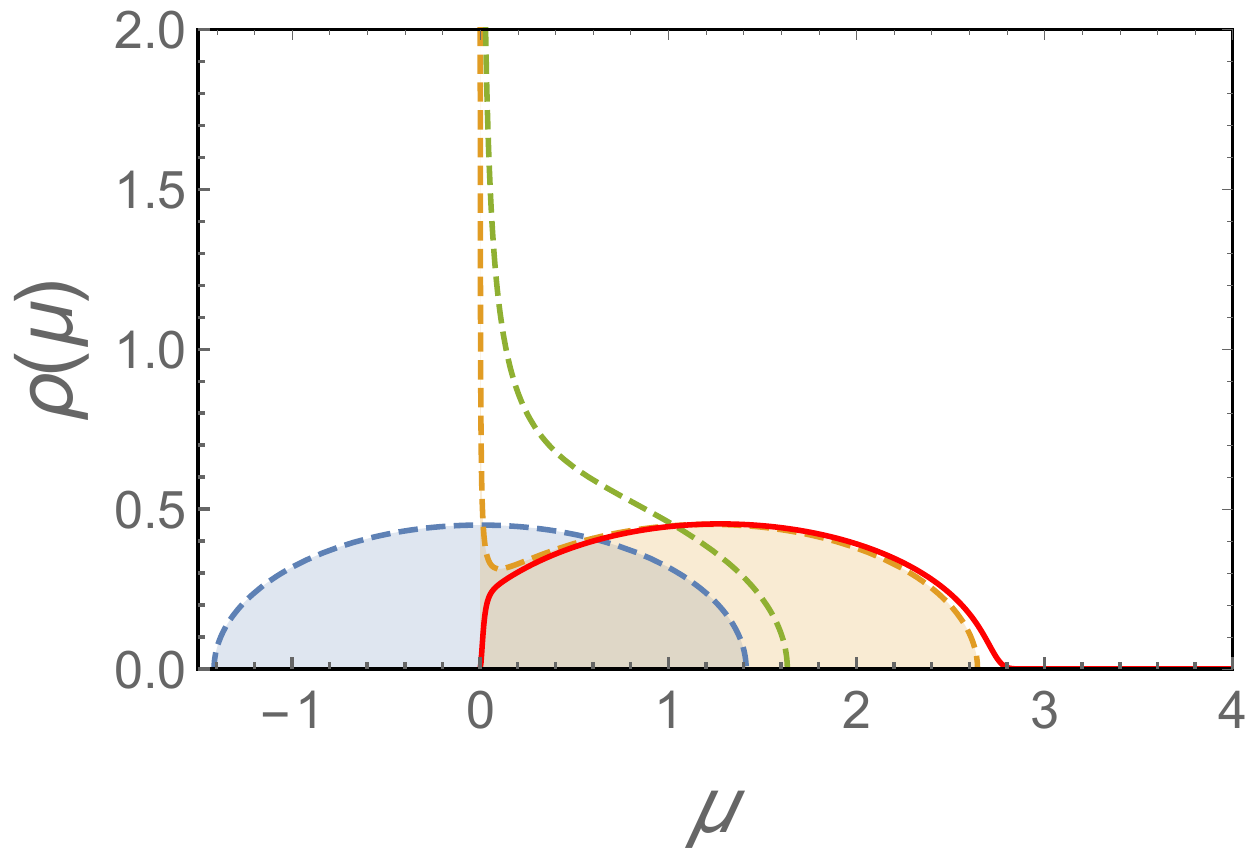} 
   \caption{
Equilibrium solution of Fokker-Planck equation for Hessian eigenvalue distribution in RGF at inflection points of the potential (solid red shaded line).  We plot also the Wigner semi-circle (dashed blue shaded line) and analytic solutions for the GOE and RGF model (dashed green and orange lines, respectively). 
      }
   \label{fig:FP-RGF3}
\end{figure}

The distribution obtained by numerically evolving the Fokker-Planck equation is shown in Fig.~\ref{fig:FP-RGF3}.  We find that the constant $C$ in Eq.~(\ref{Cmu^2}) is $\sim 0.1 N^2$ from our numerical results. As in Sec.~\ref{sec:dynamical2-2}, the average number of eigenvalues in the range $0<\mu\lesssim 1/N$ is ${\cal O}(1)$, so we cannot expect our distribution to be accurate in this range.

As before, the second smallest eigenvalue of the Hessian, $\mu_2$, 
can also be estimated from 
\beq
 \int_0^{\mu_2} \rho_c (\mu) \dd \mu \sim \frac{1}{N-1}, 
\eeq
which gives
\beq
 \mu_2 \sim \frac{1}{N}, 
 \label{mu_2}
\eeq
in agreement with (\ref{mu_2:analytic}).

The rescaled eigenvalue (\ref{mu_2}) corresponds to $\lambda_2 \sim U_0 / (\sqrt{N} \Lambda^2)$.  If this is smaller than about $H^2 \sim U_0$, then the associated field $\phi_2$ 
will undergo significant fluctuations and may play a dynamical role during inflation. 
This is unlikely if $\sqrt{N} \Lambda^2 \ll 1$.
Thus we conclude that multifield inflation is not likely for $\Lambda \ll N^{-1/4} \sim 0.3$ (for $N = 100$).

\section{Conclusions}
\label{sec:conclusion}

The main focus of this paper was to investigate the Hessian eigenvalue distribution at local minima of a random Gaussian landscape.  Bray and Dean used the saddle point approximation to calculate this distribution and the density of local minima in the leading order of the large $N$ expansion. We found, however, that the next-to-leading order corrections modify the distribution at the lower edge of the domain. This is particularly important for the smallest Hessian eigenvalues, which we need to estimate for assessing the vacuum stability and the multi-field nature of inflation in the landscape.

We extended the saddle point method to account for the sub-leading in $1/N$ contributions and used it to calculate the density of local minima in the landscape.  This method can also be used to determine the Hessian eigenvalue distribution at a generic point in the landscape, but it fails to find the distribution at potential minima with the desired accuracy.  For that we had to develop a completely new approach.

In our new approach, the Hessian eigenvalue distribution is calculated as the asymptotic endpoint of a stochastic process, called Dyson Brownian motion.  The distribution is evolved via a suitable Fokker-Planck equation, and the equilibrium distribution is obtained after a sufficiently large number of iterations.  We have verified that this method agrees with the saddle point method in cases where the latter method is applicable.

We discussed some implications of our results for vacuum stability and slow-roll inflation in the landscape.  We found that metastable vacua in a Gaussian landscape are more stable than a naive estimate would suggest.  Slow-roll inflation at inflection points in the landscape is likely to be single-field when the smallest nonzero Hessian eigenvalue at a typical inflection point is greater than the energy scale of the landscape $U_0$.  We found that this condition is satisfied if the correlation length in the landscape is $\Lambda \lesssim N^{-1/4}$.  For $N\sim 100$, this means that inflation is essentially single-field in a landscape with $\Lambda \lesssim 0.3$ in Planck units.

In Appendix A we discussed the relation between a random Gaussian landscape and an axionic landscape.  We specified the conditions under which an axionic landscape can be approximated by an isotropic random Gaussian field.  We expect that our results should be applicable to such axion models.

We note finally that the problem of Hessian eigenvalue distribution in a random field arises in many areas of condensed matter physics (see, e.g., \cite{Beenakker:1997, Guhr:1997ve, Eynard:2015aea} and references therein).  Our methods and results may be useful in these areas as well.

%%%%%%%%%%%%%%%%%%%%%%%%%%%%%%%%%%%%%%%%%%%%%%%%%%%%%%%%%%%%%%%%%%%%%%
\section*{Acknowledgments}
%%%%%%%%%%%%%%%%%%%%%%%%%%%%%%%%%%%%%%%%%%%%%%%%%%%%%%%%%%%%%%%%%%%%%%
We would like to thank Jose J. Blanco-Pillado for useful conversations and Thomas Bachlechner and Yan~V.~Fyodorov for their useful comments on the manuscript.  This work was supported in part by the National Science Foundation under grant 1518742.
%%%%%%%%%%%%%%%%%%%%%%%%%%%%%%%%%%%%%%%%%%%%%%%%%%%%%%%%%%%%%%%%%%%%%%

%%%%%%%%%%%%%%%%%%%%%%%%%%%%%%%%%%%%%%%%%%%%%%%%%%%%%%%%%%%%%%%%%%%%
\nocite{}
%%%%%%%%%%%%%%%%%%%%%%%%%%%%%%%%%%%%%%%%%%%%%%%%%%%%%%%%%%%%%%%%%%%%

\appendix

\section{Axion landscape}
\label{sec:axion}

In this Appendix we extend the argument of Ref.~\cite{Bachlechner:2017hsj} 
to show that under certain conditions the axion landscape can be approximately described by an isotropic random Gaussian field model. 

Axions develop a periodic potential due to non-perturbative effects.  (For a review of axions see, e.g.,~\cite{Kim:2008hd}.)
In general the potential has the form
\beq
U({\bm\theta})=\sum_{a=1}^P \Lambda_a^4 f_a(X_a +\delta_a) ,
\label{axionU}
\eeq
where 
\beq
X_a={\bm q}_a \cdot {\bm\theta} ,
\eeq
$\Lambda_a$ are the energy scales of non-perturbative effects, 
${\bm\theta}$ is an $N$-component vector, its components $\theta_i$ being the axion fields, ${\bm q}_a$ is a vector with integer components $q_{ai}$, and $f_a(X)$ are periodic functions with a period $2\pi$,
\beq
f_a(X+2\pi)=f_a (X).
\eeq
The phase constants $\delta_a$ are assumed to be random parameters with a flat distribution in the range from $0$ to $2\pi$, and $q_{ai}$ are independent random variables with a specified distribution $P_a(q_{ai})$.  Following Ref.~\cite{Bachlechner:2017hsj}, we shall assume for simplicity that all these distributions are identical: $P_a(q)=P(q)$ (although this can be easily generalized).

Since the functions $f_a(X)$ are periodic, they can be represented as
\beq
f_a(X)=\sum_{n=-\infty}^\infty f_{an}e^{inX},
\eeq
with $f_{a,-n}=f_{an}^*$.  In Ref.~\cite{Bachlechner:2017hsj}, they assume $f_a(X)=[1-\cos X]$, in which case $f_{a0}=1$, $f_{a,\pm 1}=1/2$, and $f_{an}=0$ for all $|n|>1$.
In what follows, we consider a generic periodic function $f_a (X)$.  For example, 
the strong dynamics of QCD results in a complicated 
periodic function for the QCD axion~\cite{DiVecchia:1980yfw} 
(see also Ref.~\cite{diCortona:2015ldu} for a recent work). 

String theory predicts the existence of a large number of axions, $N\gtrsim 100$ (e.g.,~\cite{Svrcek:2006yi}).  In Ref.~\cite{Bachlechner:2017hsj} it was shown that interesting alignment effects can arise in the axionic landscape if the number of terms in the potential (\ref{axionU}) is $N<P<2N$.  They also noted that for $P\gg N$ the potential approaches that for a random Gaussian field, as a consequence of the central limit theorem.  The statistical properties of this field depend on the choice of the distribution $P(q)$.

The integers $q_{ai}$ define a lattice in the $q$-space with a spacing $\Delta q=1$.   We shall assume that the variance of the distribution $P(q)$ is ${\bar q}^2
 \gg 1$, which means that the correlation length of $U({\bm\theta})$ is small compared to the periodicity length $2\pi$.  Then the distribution $P(q)$ can be approximated as continuous.

We will be interested in the two-point correlation function for the potential $U({\bm\theta})$.  After averaging over the random phases, this function should depend only on the difference ${\bm\theta}_1 -{\bm\theta}_2$.  Then, without loss of generality, we can choose one of the points to be at ${\bm\theta}=0$.  Thus, we consider
\beq
\left< U({\bm \theta})U(0) \right> = \sum_{a,a'} \sum_{n,n'} \Lambda_a^4 \Lambda_{a'}^4 f_{an} f_{a' n'} \left< e^{inX_a} \right>_q \left< e^{in\delta_a} e^{in'{\delta_a}'}\right>_\delta, 
\label{UU}
\eeq
where 
$\la \cdots \ra_{\alpha}$ ($\alpha = q, \delta$) 
represents the ensemble average over random variables $\alpha$. 
We use
\beq
\left<e^{in\delta_a}e^{in'\delta_{a'}}\right>_\delta =\delta_{aa'}\delta_{n+n'}
\eeq
and
\beq
\left<e^{inX_a}\right>_q = \prod_i F(n\theta_i),
\eeq
where
\beq
F(\theta)\equiv \sum_{q=-\infty}^\infty P(q) e^{i q \theta} .
\label{F}
\eeq
Substituting this in (\ref{UU}), we have
\beq
\left<U({\bm\theta})U(0)\right> -{\bar U}^2=\sum_{a,n\neq 0} \Lambda_a^8 |f_{an}|^2 \prod_i F(n\theta_i).
\eeq

Now we consider some possible forms of $P(q)$.  The first is
\beq
P(q)\propto \exp \left(-\frac{q^2}{2{\bar q}^2}\right).
\label{GaussianP(q)}
\eeq
In this case, the combined distribution for all $q_{ai}$ components is
\beq
P({\bm q}_a)\propto \exp \left(-\frac{{\bm q}_a^2}{2{\bar q}^2}\right).
\label{only}
\eeq
This depends only on ${\bm q}_a^2 =\sum_i q_{ai}^2$ and thus is rotationally invariant in the ${\bm q}$-space.  Note that Eq.~(\ref{only}) is the only factorized distribution, $P({\bm q}_a) = \prod_i P_i(q_{ai})$, that has this property.  We can expect the two-point function to also be rotationally invariant in the ${\bm\theta}$-space.  Indeed, from Eq.~(\ref{F}) we have
\beq
F(\theta) \propto \sum_{q=-\infty}^\infty \exp\left(-\frac{q^2}{2{\bar q}^2} +iq\theta\right).
\eeq

Since we assume that ${\bar q}\gg 1$, the sum over $q$ can be approximated by an integral, so we obtain
\beq
F(\theta) \propto \exp\left(-\frac{1}{2}{\bar q}^2 \theta^2\right)
\eeq
and
\beq
\left<U({\bm \theta})U(0)\right>-{\bar U}^2 \propto\sum_{a,n} \Lambda_a^8 |f_{an}|^2 \exp\left(-\frac{1}{2}{\bar q}^2 n^2 {\bm\theta}^2\right).
\label{theta^2}
\eeq
Note that if $\bar{q}$ depends on the index $a$, $\bar{q}$ in this equation should simply be replaced by $\bar{q}_a$.

Now let us consider the form of $P(q)$, which was adopted in Ref.~\cite{Bachlechner:2017hsj}: $P(q) = {\rm const}$ for $|q|<q_{\rm m}$ and $P(q)=0$ otherwise.  In this case,
\beq
F(\theta)\propto \frac{\sin(q_{\rm m}\theta)}{\theta}.
\eeq
Hence,
\beq
\left<U({\bm\theta})U(0)\right>-{\bar U}^2 \propto \prod_i \frac{\sin(q_{\rm m}n\theta_i)}{\theta_i}.
\eeq
Unlike Eq.~(\ref{theta^2}), this correlation function is not rotationally invariant.  The reason is that rotational invariance is violated by the probability distribution for ${\bm q}$.

It may be instructive to compare the correlators
of the potential $U({\bm\theta})$ and its derivatives for different choices of $P(q)$. 
We find 
\beq
 &&\bar{U} \equiv \la U ({\bm \theta} ) \ra_{q, \delta}  
 = \sum_a \Lambda_a^4 \la f_a \ra_\delta, 
 \\
 &&\la (U ({\bm \theta} )  - \bar{U})^2 \ra_{q, \delta}   
 = \sum_a \Lambda_a^8  \la \lmk f_a - \la f_a \ra \rmk^2 \ra_\delta, 
 \\
 &&\la U ({\bm \theta}) \zeta_{ij} ({\bm \theta}) \ra_{q, \delta}  
 = \sum_a \Lambda_a^8 \la q_{ai} q_{aj} \ra_q \la f_a f''_a \ra_\delta, 
 \\
 &&\la \zeta_{ij} ({\bm \theta}) \zeta_{kl} ({\bm \theta}) \ra_{q, \delta}  
 = 
 \sum_a \Lambda_a^8 \la q_{ai} q_{aj} q_{ak} q_{al} \ra_q \la f_a''^2 \ra_\delta, 
 \label{correlator-zeta}
\eeq
where primes denote derivatives with respect to $X$. 
The gradient $\eta_i = \partial U/\partial\theta_i$ is not correlated with the potential nor the Hessian.
The ensemble average over $q_{ai}$ gives 
\beq
 &&\la q_{ai} q_{aj} \ra_q 
 = 
 \delta_{ij} \bar{q}_a^2 
 \\
 &&\la q_{ai} q_{aj} q_{ak} q_{al} \ra_q
 = \bar{q}_a^4 \lmk \delta_{ij} \delta_{kl} + \delta_{ik} \delta_{jl} + \delta_{il} \delta_{jk} 
 - r_a \delta_{ij} \delta_{jk} \delta_{kl} \rmk
\label{A22}
 \\
 && r_a \equiv \frac{3 \bar{q}_a^4 - \la q_{ai}^4 \ra_q}{\bar{q}_a^4}, 
\eeq
where $\bar{q}_a^2$ is the variance of random variable $q_{ai}$.

If we identify 
\beq
 &&E = \sum_a \Lambda_a^8  \la \lmk f_a - \la f_a \ra \rmk^2 \ra_\delta
 \\
 &&B = \sum_a \Lambda_a^8 \bar{q}_a^2 \la f_a f''_a \ra_\delta
 \\
 &&A = 
 \sum_a \Lambda_a^8 \bar{q}_a^4 \la f_a''^2 \ra_\delta, 
\eeq
we see that the resulting correlation functions have the same form as Eqs.~(\ref{1})-(\ref{2}),  
except for the additional term ($r \delta_{ij} \delta_{jk} \delta_{kl}$) in (\ref{A22}). 
This additional term breaks the rotational invariance of the model and vanishes when 
the random variables $q_{ai}$ have a rotationally invariant distribution (\ref{only}).
In Ref.~\cite{Bachlechner:2017hsj}, the authors assumed 
a flat distribution for $q_{ai}$ as an example (which breaks rotational invariance) 
and obtained $r = 6/5$.

The extra term in (\ref{A22}) also indicates a deviation from Gaussian statistics.  However, 
it affects only the statistics of the diagonal components of the Hessian. There are only $N$ diagonal components and $N(N-1)/2$ non-diagonal ones, so we can expect this term to be unimportant at large $N$.  The same discussion applies to correlators for higher-order derivatives. Thus we expect that 
Gaussian random fields give a good approximation for this type of landscape in 
the limit of $N \gg1$ and $P \gg N$.

Finally, we comment that 
the cancellation for the coefficient of $(\Tr \zeta)^2$ ($AE - B^2 = 0$) 
occurs when 
$f_a (X) = [1-\cos X ]$, 
which was adopted in Ref.~\cite{Bachlechner:2017hsj}. 
In this case, the Hessian distribution is just given by the GOE with a constant shift 
of the diagonal terms as \eq{shifted GOE}. 
However, this is not a generic property of axion landscape with a generic choice of functions $f_a(X)$.

\end{document}